\documentclass[12pt,preprint]{aastex} 

\shortauthors{Oliveira et al.}
\shorttitle{New Young Stellar Population in Serpens}

\begin{document}

\title{Optical Characterization of a New Young Stellar Population in the Serpens Molecular Cloud}
\author{ Isa Oliveira\altaffilmark{1,2}, Bruno Mer\'{\i}n\altaffilmark{3,2}, Klaus M. Pontoppidan\altaffilmark{1,4}, Ewine F. van Dishoeck\altaffilmark{2,5}, Roderik A. Overzier\altaffilmark{6}, Jes\'us Hern\'andez\altaffilmark{7,8}, Aurora Sicilia-Aguilar\altaffilmark{9}, Carlos Eiroa\altaffilmark{10},  Benjamin Montesinos\altaffilmark{11,12}}
\altaffiltext{1}{California Institute of Technology, Division for Geological and Planetary Sciences, MS 150-21, Pasadena, CA 91125, USA; email: isa@gps.caltech.edu}
\altaffiltext{2}{Sterrewacht Leiden, Leiden University, P.O. Box 9513, 2300 RA  Leiden, The Netherlands}
\altaffiltext{3}{Research and Scientific Support Department (ESTEC/ESA), PO Box 299, 2200 AG Noordwijk, The Netherlands}
\altaffiltext{4}{Hubble Fellow}
\altaffiltext{5}{Max-Planck-Institut f\"ur Extraterrestriche Physik, P.O. Box 1312, D-85741 Garching, Germany}
\altaffiltext{6}{Max-Planck-Institut f\"ur Astrophysik, Karl-Schwarschild-Str. 1, D-85741 Garching, Germany}
\altaffiltext{7}{Department of Astronomy, University of Michigan; Ann Arbor, MI 48109, USA}
\altaffiltext{8}{Centro de Investigaciones de Astronom\'{\i}a, M\'erida 5101-A, Venezuela}
\altaffiltext{9}{Max-Planck-Institut f\"ur Astronomie, K\"onigstuhl 17, 69117 Heidelberg, Germany}
\altaffiltext{10}{Universidad Aut\'onoma de Madrid, Departamento de F\'{\i}sica Te\'orica C-XI, 28049 Madrid, Spain}
\altaffiltext{11}{Instituto de Astrof\'{\i}sica de Andaluc\'{\i}a, Apartado 3004, E-18080 Granada, Spain}
\altaffiltext{12}{Laboratorio de Astrof\'{\i}sica Espacial y Fisica Fundamental, ESAC, Apartado 78, E-28691 Villanueva de la Ca\~nada, Madrid, Spain}

\begin{abstract}

We report on the results of an optical spectroscopic survey designed to confirm the youth and determine the spectral types among a sample of young stellar object (YSO) candidates in the Serpens Molecular Cloud. We observed 150 infrared excess objects, previously discovered by the Spitzer Legacy Program ``From Molecular Cores to Planet-Forming Disks'' (c2d), bright enough for subsequent Spitzer/IRS spectroscopy. We obtained 78 optical spectra of sufficient $S/N$ for analysis. Extinctions, effective temperatures and luminosities are estimated for this sample, and used to construct H-R diagrams for the population. We identified 20 background giants contaminating the sample, based on their relatively high extinction, position in the H-R diagram, the lack of H$\alpha$ emission and relatively low infrared excess. Such strong background contamination (25\%) is consistent with the location of Serpens being close to the Galactic plane ($5^{\circ}$ Galactic latitude). The remaining 58 stars (75\%) were all confirmed to be young, mostly K and M-type stars that are presumed to belong to the cloud. Individual ages and masses for the YSOs are inferred based on theoretical evolutionary models. The models indicate a spread in stellar ages from 1 to 15 Myr, peaking at 2 -- 6 Myr, and a mass distribution of 0.2 to 1.2 $M_\odot$ with median value around 0.8 $M_\odot$. Strong H$\alpha$ emission lines (EW[H$\alpha$] $>$ 3 \AA) have been detected in more than half of the sample (35 stars). The mass accretion rates as derived from the H$\alpha$ line widths span a broad distribution over 4 orders of magnitude with median accretion rate of $10^{-8}$ $M_\odot$ yr$^{-1}$. Our analysis shows that the majority of the infrared excess objects detected in Serpens are actively accreting, young T-Tauri stars.

\end{abstract}


\section{Introduction}
\label{sintro}

\noindent
The Serpens molecular cloud has received considerable attention over the past decade. Because it is an actively star-forming complex containing a substantial mass of molecular gas and young stars within both clustered and diffuse environments, it has become one of our main laboratories for testing theories of low-mass star formation (\citealt{DJ06}; \citealt{EI05}; \citealt{KL04}; \citealt{KA04}; \citealt{PR03}; \citealt{OT02}; \citealt{WM00}; \citealt{HO99}; \citealt{PR98}; \citealt{HE97}).

The Serpens cloud is one of the five clouds selected as part of the Spitzer Legacy program ``From Molecular Cores to Planet-Forming Disks'' (c2d; \citealt{EV03}), providing images in the 3 -- 70 $\mu$m range. The wide wavelength coverage and high sensitivity in the infrared (IR) of Spitzer make it possible to easily identify a new, complete, flux limited (down to luminosities below 0.01 $L_\sun$) young stellar population. This population has been found to be distributed over almost the entire area surveyed, offering an opportunity to determine the stellar content in different regions of the cloud, the distributions of the youngest stars and substellar objects, and the properties of their circumstellar envelopes and disks.

The c2d program has mapped a 0.89 deg$^2$ portion out of the more than 10 deg$^2$ area \citep{KA04} of the Serpens molecular cloud. Assuming a distance of 259 $\pm$ 37 pc \citep{ST96}, this corresponds to a covered area of about 2.5 $\times$ 9 pc. This region was discovered to be very rich in young stars (\citealt{HA07}), some of them previously identified with ISO data (\citealt{DJ06}). This poorly known cluster of young stars, cluster B, and the previously unknown star-forming region around it, is located about half a degree southwest of the well-known Serpens Cloud Core, containing cluster A \citep{KA04}. It has a high density of young stars, making it an unique target region for obtaining a full and consistent picture of clustered low-mass star formation and compare this with young stars in the surrounding field. 

We are currently carrying out a multi-wavelength survey of this region, from X-ray to millimeter wavelengths, in order to create a ``template'' sample for the study of the evolution of circumstellar disks around stars younger than $\sim10$ Myr, within a single, small and well defined region. This work provides the necessary information for anchoring the study of protoplanetary disks to their parent population by means of the optical spectroscopic classification of the central stars. Precise stellar properties are needed to perform an accurate study of the evolutionary stages of the young stellar objects and their disks. Most spectroscopic studies of protoplanetary disk evolution refer to samples of young stars scattered across the sky or to sources in large star-forming clouds like Taurus, making it difficult to separate intrinsic evolutionary effects from those caused by external influences such as environment or star formation history.

In this paper we report on our optical spectroscopic survey designed to confirm the youth and determine spectral types of the newly discovered young stellar object (YSO) candidates in the Serpens region with Spitzer. These spectra will also be used to estimate mass accretion rates using the strength of the H$\alpha$ emission line (\citealt{MU03}; \citealt{WB03}, \citealt{NA04}), and the extinction toward each object.

The paper is constructed as follows: \S~\ref{ssample} describes the selection criteria for our sample, and in \S~\ref{sobs} we describe the observations and data reduction. In \S~\ref{sspectype} we discuss our spectral classification methods, along with its results for this sample and derive effective temperatures. In \S~\ref{sextin}, we present our extinction studies based on the Spitzer photometry and on our optical spectra. In \S~\ref{shrd} we derive luminosities for the objects. This allows us to place each object in an Hertzsprung-Russell (HR) diagram, and derive its age and mass from comparison with evolutionary models. The H$\alpha$ line and accretion rates are discussed in \S~\ref{sacret}. Finally, in \S~\ref{scon} we present our conclusions.

\section{Sample Selection}
\label{ssample}

\noindent
From the c2d catalog, \citet{HA07} plotted spectral energy distributions (assuming a K7 spectral type for all the sources) and identified 235 YSO candidates -- objects with IR excess, interpreted as due to disks or envelopes. These authors discuss the problem that background galaxies and bright background post-AGB stars may show the same color distribution as that of young stellar objects. Particularly, the post-AGB stars could contaminate the bright end of the luminosity distribution.

To this original sample of 235 objects, a flux threshold of 3mJy at 8$\mu$m was applied, resulting in 150 sources bright enough to be observed in follow-up observations with the InfraRed Spectrograph (IRS), onboard Spitzer. The IRS observations, carried out to study the evolution of dust in protoplanetary disks of a complete, flux limited sample will be presented in a future paper \citep{OL09}. Here, we present optical spectroscopy of 78 stars bright enough to obtain optical spectroscopy with a 4m-class telescope. The remaining objects are too extincted in the optical to be detected, and are being classified in an ongoing near-IR spectroscopic survey. 

The resulting sample contains objects which span a range in evolutionary stages from embedded Class I (i.e., spectral slope $\alpha_{2-24\mu{\rm m}} >$ 0.3) sources to young stars with protoplanetary disks (Class II sources, -1.6 $< \alpha_{2-24\mu{\rm m}} < $ 0.3) and stars with photospheric fluxes in the IRAC bands and excess in the MIPS bands (also called ``cold disks'', e.g. \citealt{JB07}).

\begin{figure}[h!]
\begin{center}
\includegraphics[width=0.45\textwidth]{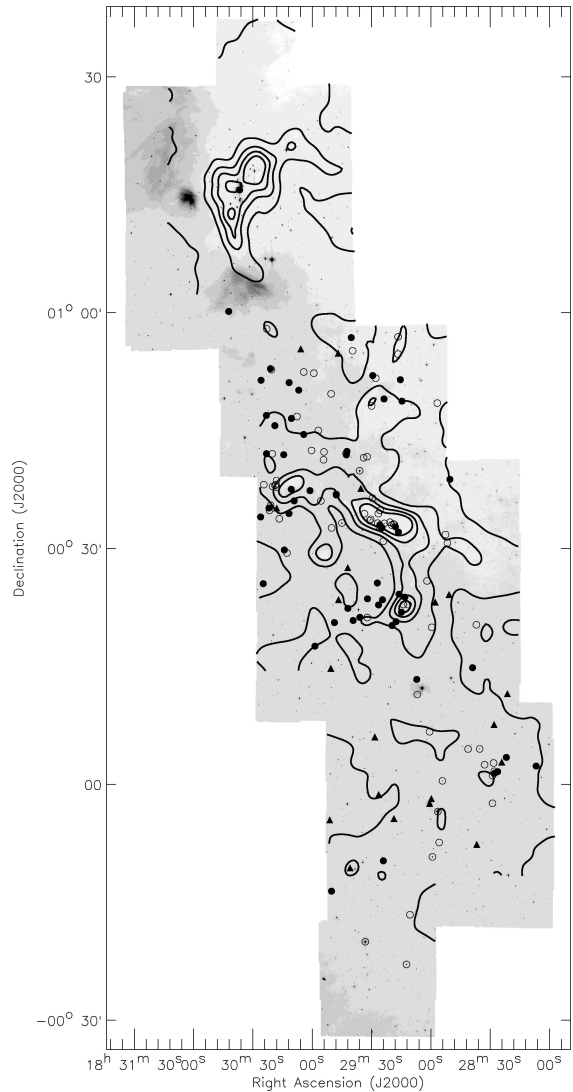}
\end{center}
\caption{\label{objirac} Observed objects shown on the 8.0$\mu$m IRAC map of the Serpens Molecular Cloud, imaged as part of the Spitzer Legacy Program ``From Molecular Clouds to Planet-Forming Disks'' (c2d) \citep{HA06}. Filled symbols are the 78 objects for which we present optical spectroscopy in this paper: filled circles are confirmed members of the cloud and filled triangles are likely candidate background contaminants (\S~\ref{shrd}). The open symbols are the remaining young stellar object candidates in the sample, not detected in our survey of 150 YSOs. Overlaid are the contours (5, 10, 15, 20 and 25 mag) of visual extinction derived from the ``c2d'' extinction maps.}
\end{figure}

Figure \ref{objirac} shows the spatial distribution of the complete sample of 150 Serpens YSOs observed with Spitzer/IRS. The filled and open symbols correspond to detected and undetected objects from the optical survey presented in this paper, respectively.

\section{Observations and Data Reduction}
\label{sobs}

\begin{figure*}[ht!]
\begin{center}
\includegraphics[width=0.8\textwidth]{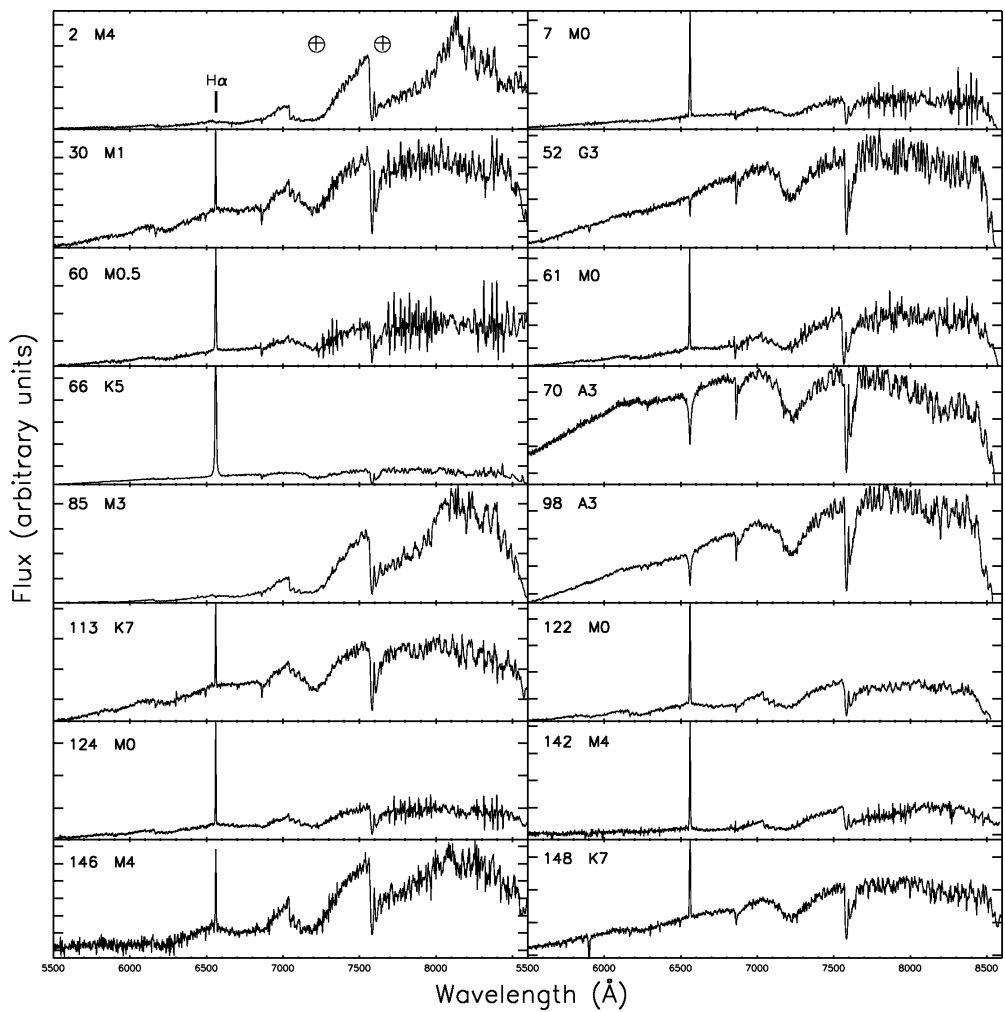} 
\end{center}
\caption{\label{spectra} Selected spectra of a representative sample of the observed objects. Along with their number from Table \ref{t_obs}, the extracted spectral types are shown for reference. The strong H$\alpha$ lines are noticeable in several spectra. The main remaining atmospheric features are indicated with $\earth$.}
\end{figure*}

\noindent
The spectroscopic data were obtained during observing runs at three telescopes: two runs with the multi-object Wide Field Fibre Optical Spectrograph ({\it WYFFOS}) at the 4.2m William Herschel Telescope ({\it WHT}) in 2006 May and 2007 June, and one run with the 3.6m Device Optimized for the LOw RESolution ({\it DOLORES}) at the Telescopio Nazionale Galileo ({\it TNG}) in 2006 June, both telescopes are located at the Observatorio del Roque de los Muchachos in La Palma, and one run with the Calar Alto Faint Object Spectrograph ({\it CAFOS}) at the 2.2m Calar Alto Telescope. No absolute flux calibration was performed, nor were telluric absorption features removed. Technical information on each run is specified in Table \ref{t_obs}. Figure \ref{spectra} shows a representative sample of the spectra of the objects in our sample.

\subsection{WHT data}
\label{whtdata}

\noindent
{\it WYFFOS} was fed from the AutoFib 2 robotic fibre positioner \citep{PA94} and targets were selected to maximize the number of objects observed in each configuration, using its program $\mathtt{AF2\_CONFIGURE}$. We used the ``small'' fibre module of {\it WYFFOS}, corresponding to fibre aperture of 1.6\arcsec. The wavelength range was $\sim$3000 \AA{}, centered at 7000 \AA, in order to include the H$\alpha$ and Li I (6707 \AA) lines. The average resolving power was $R \sim 1750$. In order to adequately remove the effects of cosmic ray hits on the detector, each field was observed using three exposures. Flat-fields and bias frames were obtained at the beginning of each night and neon calibration arc lamp spectra at the beginning of the nights, as well as separately for each configuration. Unused fibres were placed on the sky. We obtained {\it WYFFOS} spectra for 71 objects of our sample. 

Data reduction was performed using our own code, based on the instrument reduction guideline\footnote{http://www.ing.iac.es/Astronomy/instruments/af2/index.html}, within $\mathtt{IRAF}$ and $\mathtt{IDL}$. After bias subtraction and flat-fielding, the spectra were extracted and wavelength calibrated with the arc Ne comparison lamp data, to a precision of $\sim$0.5 \AA.

\subsubsection{Sky subtraction}

\noindent
The quality of the optical spectra is limited by the accuracy of the sky subtraction, especially for faint stars. As opposed to long-slit spectroscopy, where an adequate sky removal can be performed using the sky background on opposite sides very close to the object spectrum, a precise sky subtraction in multi-object fibre spectroscopy is complicated by the lack of a sky spectrum close to the object spectrum both in position and in time (e.g., see \citealt{WY92}). In general, the sky background depends mainly on the local sky conditions at the observatory during the time of observations. However, in the case of a molecular cloud, the ``sky'' may also have a significant contribution from extended continuum and line emission from the cloud itself.

We used the fibres placed on ``empty'' positions in the field to measure the contribution from the sky. The sky spectra were reduced and extracted in an identical manner to the science data. The sky spectra were visually inspected and a few of bad quality were discarded. None of them showed diffuse H$\alpha$ in emission from the cloud. To reach optimum results, even for the faintest sources, a single master sky spectrum, built with an average of 12 -- 15 sky spectra, was computed for each configuration. During the sky subtraction of each star, we applied small scalings to the master sky spectrum to take into account the fact that the throughput of different fibres shows slight variations. The scalings were typically in the range 0.95--1.05, and chosen to optimize the sky subtraction. 

In several cases, the sky subtraction proved imperfect with residuals being most notable at $\lambda > $7700\AA{}. These residuals are very common in low to intermediate $S/N$ fibre spectra due to the strong OH sky lines beyond 6700 \AA. It is very difficult to remove these OH features because the sky lines are usually not well resolved, and also due to the non-linear variations in throughput and dispersion for different fibres. We have checked, however, that the spectral type classification and H$\alpha$ equivalent width measurements are not substantially hampered by these effects.

\subsection{TNG data}
\label{tngdata}

\noindent
The observations with {\it DOLORES} used the long-slit mode. Slit rotation was calculated in order to enable more than one object in the slit whenever two targets were closer than the length of the slit (8\arcmin{}). The wavelength range of $\sim$1600 \AA\ was centered at 7000 \AA, again including the H${\alpha}$ and Li I (6707 \AA) lines, with resolving power $R \sim 2500$. Exposure times ranged from 12 to 1000 seconds. Flat-fields, bias and argon comparison arc lamp spectra were obtained at the beginning of each night.

After bias subtraction and flat-fielding, the spectra were extracted and wavelength calibrated with a combination of the Ar arc lamp data and airglow lines\footnote{http://astro.u-strasbg.fr/$\sim$koppen/divers/AirGlow.html}, including the OH Meinel bands, using the ``doslit'' task of the $\mathtt{specred}$ package in $\mathtt{IRAF}$. Sky subtraction was done during the extraction based on the off-source regions along the slit. In total, we obtained {\it DOLORES} spectra of 18 sources, 15 of which were also observed with {\it WYFFOS}.

\subsection{CAFOS data}
\label{cafosdata}

\noindent
{\it CAFOS} was used with mask observations in 2008 June. Three masks were taken using the grism R-100. Flat-fields, bias and HgCd/He/Rb comparison arc lamp spectra were obtained at the beginning of each night. The wavelength coverage (3500 \AA{}, centered on 7750 \AA{}) for each source varies slightly depending on their position on the CCD, with resolving power $R \sim 3523$. The exposure time of 2400 seconds for each mask was divided in two for cosmic ray rejection.

After bias subtraction and flat-fielding, the fields were divided so that each source would be in an image by itself and then further reduced as standard slit observations: the spectra were extracted and wavelength calibrated using the ``doslit'' task of the $\mathtt{specred}$ package in $\mathtt{IRAF}$, performing sky subtraction during the extraction based on the off-source regions along each slit. 

The low resolution of these instruments, combined with low S/N for most faint objects, did not allow us to detect or put meaningful upper limits on the Li I (6707 \AA) line.

\section{Spectral Classification}
\label{sspectype}

Two methods were used in the spectral classification of the sources and are explained below.

\subsection{Method I}
\label{method_01}

\begin{figure}[h!]
\begin{center}
\includegraphics[width=0.45\textwidth]{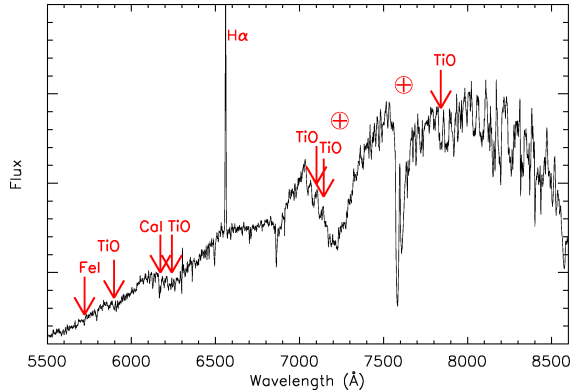} 
\end{center}
\caption{\label{feat} Features used in the spectral type classification scheme of \citet{JE04} are shown on the spectrum of source \# 113. }
\end{figure}

\noindent
The objects were classified following the spectral classification scheme of \citet{JE04} optimized to classify early type stars (up to early G). For stars with spectral type later than G, the scheme was extended by incorporating spectral indices for the ranges G0-K5 and K5-M6 (\citealt{SI05}, \citealt{BR05}\footnote{Information about the code at http://www.astro.lsa.umich.edu/$\sim$hernandj/SPTclass/sptclass.html}). More information about this classification scheme can be found in the references above. In short, it is based on strong spectral features that are sensitive to $T_{{\rm eff}}$ but insensitive to reddening in a specific luminosity class (dwarf in this case), since it uses local continua for the determination of the feature's equivalent widths (EW). It is also designed to avoid problems caused by non-photospheric contributions (e.g. emission lines). The wavelength range of the {\it WYFFOS} spectra covers 12 of these features for all spectral types, while {\it DOLORES} covers 6 and the range of {\it CAFOS} covers 15 features. On average, 6 features for {\it WYFFOS}, 5 for {\it DOLORES} and 7 features for {\it CAFOS} were used for each classification, as exemplified in Figure \ref{feat}.

The spectral types calculated from each feature result in a weighted mean spectral type. The errors in the estimated spectral types have two contributions: the error in the measurement of each feature and the error of its fit to the standard (see \citealt{JE04} for more details). A few objects were observed in both runs (as can be seen in Table \ref{t_obs}). For these objects, both spectra were ran separately through the spectral classification code. Their spectral classifications were found to be consistent with each other, typically within one subclass.

\subsection{Method II}
\label{method_02}

\noindent
Due to telluric absorption features not corrected for in the spectra presented here, care needs to be taken when using automatic fitting of features very close to these telluric bands. To make sure that the spectral types derived in Method I are correct, an additional spectral typing procedure was applied to the data.

\begin{figure}[h!]
\begin{center}
\includegraphics[width=0.7\textwidth]{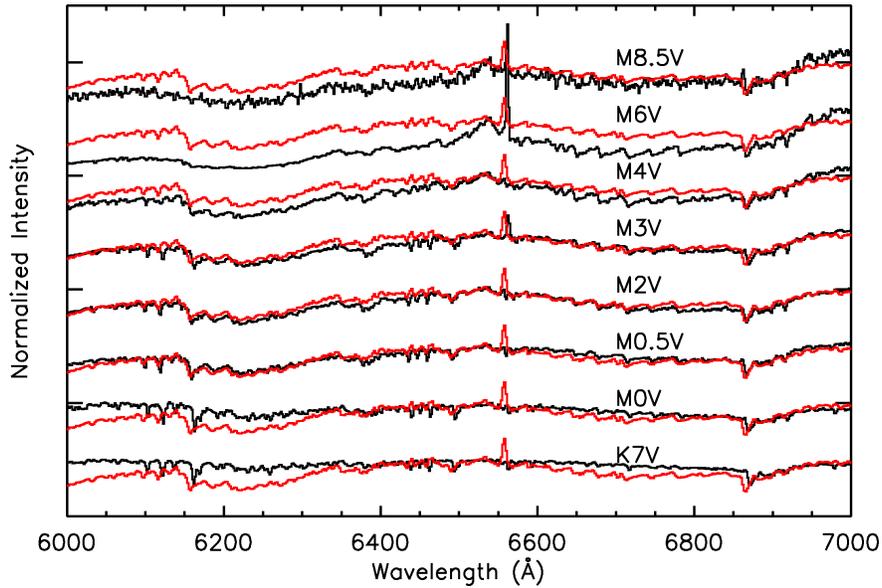} 
\end{center}
\caption{\label{feat2} Correlation between science spectrum of source \#115 (red) and standard spectra for different spectral types (black; spectral types indicated on the top right of each standard spectrum from the library of \citet{MO01}). }
\end{figure}

The library of standard spectra from \citet{MO01}, covering the wavelength range of 5800--7000\AA{}, was used in this method. In this region, different lines and bands are found for specific spectral types, making them unique. Each science spectrum was normalized and compared to normalized standards of different spectral types. The spectral classification procedure was based on finding the best correlation between observed and standard spectra, combined with a visual inspection of each match. Figure \ref{feat2} shows an example of the correlation.

\subsection{Results}
\label{method_03}

\begin{figure}[h!]
\begin{center}
\includegraphics[width=0.7\textwidth]{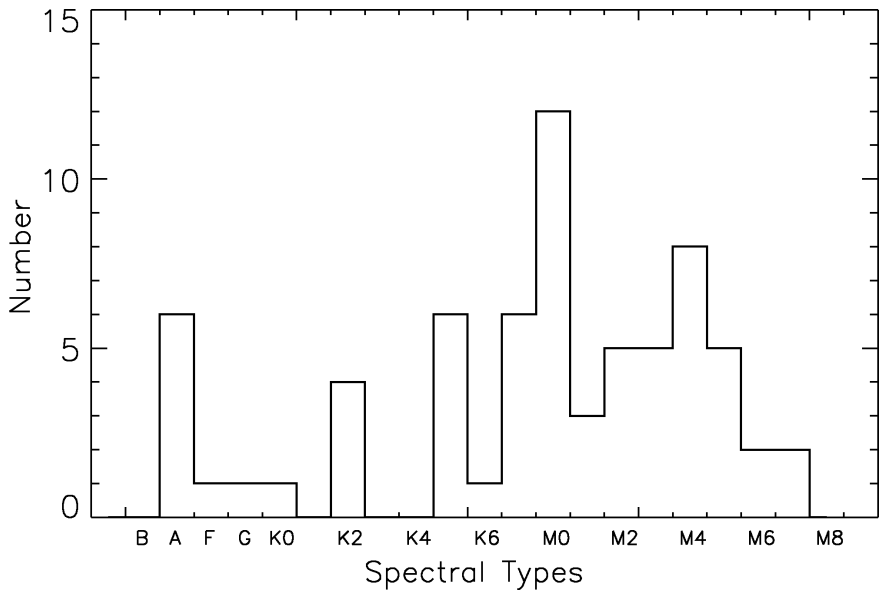} 
\end{center}
\caption{\label{hist} Distribution of spectral types of YSO candidates in the Serpens Molecular Cloud based on the classification scheme described in \S~\ref{sspectype}. } 
\end{figure}

\noindent
The two methods agree in their classification (within 1 subclass) for about 80\% of the sources. When in disagreement, Method II is judged to be more reliable because of the visual inspections, whereas Method I may be contaminated by the forementioned telluric absorption features. The spectral types obtained for our sample are given in Table \ref{t_id}. Figure \ref{hist} shows the distribution of spectral types, which is clearly dominated by K and M-type stars.

\subsection{Effective Temperature}
\label{steff}

\noindent
Effective temperatures ($T_{{\rm eff}}$) were determined using calibrations relating them to spectral types. For stars earlier than M0, the relationship established by \citet{KH95} was adopted, while for later type stars that by \citet{LU03} was used. This method was also used in the determination of $T_{{\rm eff}}$ for similar study on another of the clouds observed by the c2d, Cha II \citep{SP08}. The errors in $T_{{\rm eff}}$ come directly from the errors in the spectral type determination.

\section{Visual Extinction}
\label{sextin}

\noindent
Once the spectral types are known, we used synthetic NEXTGEN spectra \citep{HA99} and the extinction law by \citet{WD01} to compute the observed extinction towards our targets ($A_V$Obs). The appropriate synthetic spectrum for each spectral type was extincted with $A_V$ ranging from 0 to 20 mag (with increments of 0.5 mag) and a $\chi^2$ minimization was performed between the observed and synthetic extincted spectra to estimate the extinction. We estimate an intrinsic error in our $A_V$Obs determinations of 1.0-1.5 mag by comparing our source 98 (spectral type A3) with the A3 V absolute flux standard star Kopff 27 \citep{ST77}, from the ING list of spectro-photometric standard stars. This error includes the uncertainty in the total extinction determination and the fact that the fiber-fed spectra are not flux calibrated.

Table \ref{t_id} compares our extinction values from the optical spectra ($A_V$Obs) with those from the ``c2d'' extinction maps ($A_V$Cloud). The ``c2d'' map was constructed using the average extinction toward background stars within a 5\arcmin{} beam. Thus, these values give an indication of the average amount of extinction that the cloud produces in a given area, albeit with a large beam. More information about this extinction determination and its uncertainties can be found in the ``c2d'' delivery documentation \citep{EV06b}, and in \citet{HA06}.

Figure \ref{extinction} compares the two estimates of extinction for our targets described above. A large spread around the line of equal extinction is seen. This dispersion is useful to pinpoint objects with peculiar extinction characteristics. Taking into account a typical error of 2 mag in both extinction determinations and the large beam sizes for the ``c2d'' extinction values, the larger differences might be explained by two effects:

\begin{figure}[h!]
\begin{center}
\includegraphics[width=0.6\textwidth]{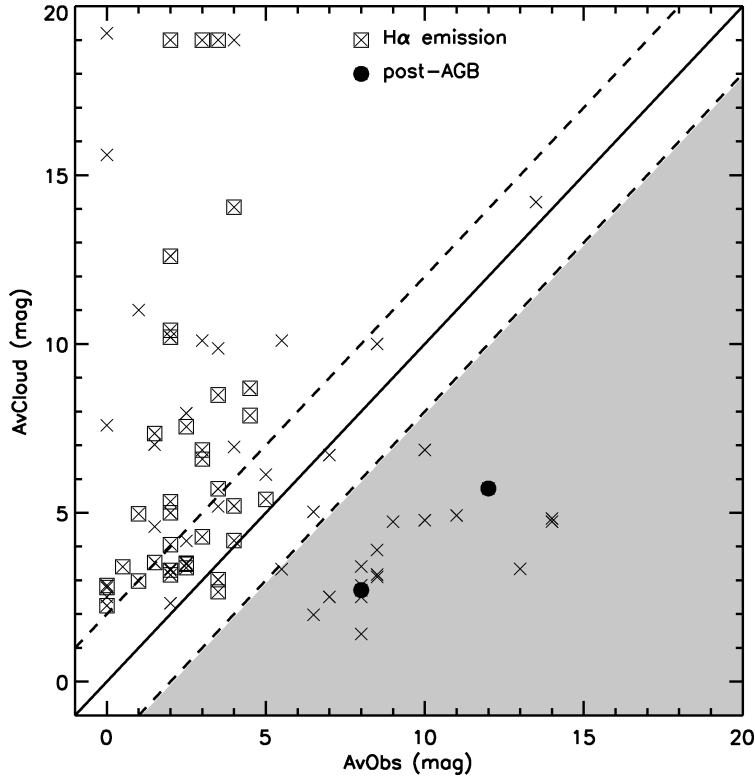}
\end{center}
\caption{\label{extinction} The relationship between the extinction values estimated from our optical spectra ($A_V$Obs) and the extinction values from the ``c2d'' Spitzer data averaged over a 5$'$ radius ($A_V$Cloud) is shown with crosses. The solid line indicates $A_V$Obs = $A_V$Cloud, with typical errors of 2 magnitudes (dashed lines). Objects marked with a square are accreting stars according to the definition of \citet{WB03}, while objects marked with a filled circle are confirmed post-AGB stars. The shaded region is where we propose background stars should lie. }
\end{figure}

(i) For the sources with $A_V$Obs $\lesssim A_V$Cloud, an indication of their depth in the cloud can be obtained. Specifically, such sources are likely embedded in the cloud, or located in front of it.

(ii) If $A_V$Obs $>> A_V$Cloud (shaded area in Figure \ref{extinction}), the molecular cloud by itself is unlikely to explain such extinction. One one hand, for young stars embedded in their parent envelopes, an extra extinction compared to the local average in the cloud is naturally expected. Such embedded sources, however, are also expected to show a strong infrared excess. On the other hand, stars at distances much larger than that of the cloud are also likely to have an enhanced extinction. We argue that this is the dominant category of sources in the shaded area of Figure \ref{extinction}, though confirmation through another method is needed. 

Indeed, none of the objects in the shaded area was found to have a strong infrared excess, as evidenced by the low Spitzer-derived slope ($\alpha_{2-24\mu{\rm m}}$) values, shown in Table \ref{t_id}. This suggests that those objects are not surrounded by large amounts of circumstellar dust but rather by tenuous dust layers, consistent with them being post-AGB stars at distances of several kpc. Follow-up c2d observations of bright YSO candidates in the Serpens cloud with Spitzer IRS confirmed that two of the sources with $A_V$Obs $>> A_V$Cloud are post-AGB stars, seen through the cloud (filled circles in Figure \ref{extinction}). Post-AGB stars are so bright that, at a farther distance, their colors and magnitudes can be comparable to those of YSOs in the cloud. They can, however, be distinguished by the presence of spectral features at 13 and 21 $\mu$m, not seen in YSOs \citep{WA99}.

\section{H-R Diagram}
\label{shrd}

\noindent
Given the spectral type, therefore effective temperature (derived in \S~\ref{steff}), and luminosity of a source, individual age and mass can be estimated by comparing its position in the Hertzsprung-Russell (HR) diagram with evolutionary tracks.

\subsection{Luminosities}
\label{slum}

\noindent
Stellar luminosities were calculated by integrating the NEXTGEN stellar photosphere for each object, scaled to dereddened optical (from literature) or 2MASS fluxes, depending on their availability (see Figure \ref{sed_comp} for two examples). The errors are derived from the uncertainty in the distance to the cloud and the error on the extinction ($\pm$2 mag). Similar methods for luminosity estimation have been widely used in the literature (\citealt{VA97}, \citealt{VA98}, \citealt{AL08}, \citealt{ME08}). The extinction law of \cite{WD01} was used to correct for the reddening, to be consistent with the analysis in \S~\ref{sextin}. A careful inspection of all the resulting SEDs was performed to find the best-fit extinction from the two $A_V$ values discussed above plus the best-fit $A_V$ from the spectral type and the optical or near-IR photometry. The resulting best-fit extinctions ($A_V$Final) and luminosities are shown in Table \ref{t_hrd}. Objects \# 44 and 47 were not detected by 2MASS and, therefore, no $J$, $H$ and $K$ magnitudes are available for them. For this reason, we could not determine their luminosities and they are not included in the H-R diagram. Objects \# 41 and 117 are rising sources (likely due to an edge-on disk or envelope), resulting in an unrealistically low luminosity and  were marked with a different symbol (square) in the H-R diagram.

\subsection{Results}
\label{sres_hrd}

\begin{figure*}[ht!]
\begin{center}
\includegraphics[width=0.9\textwidth]{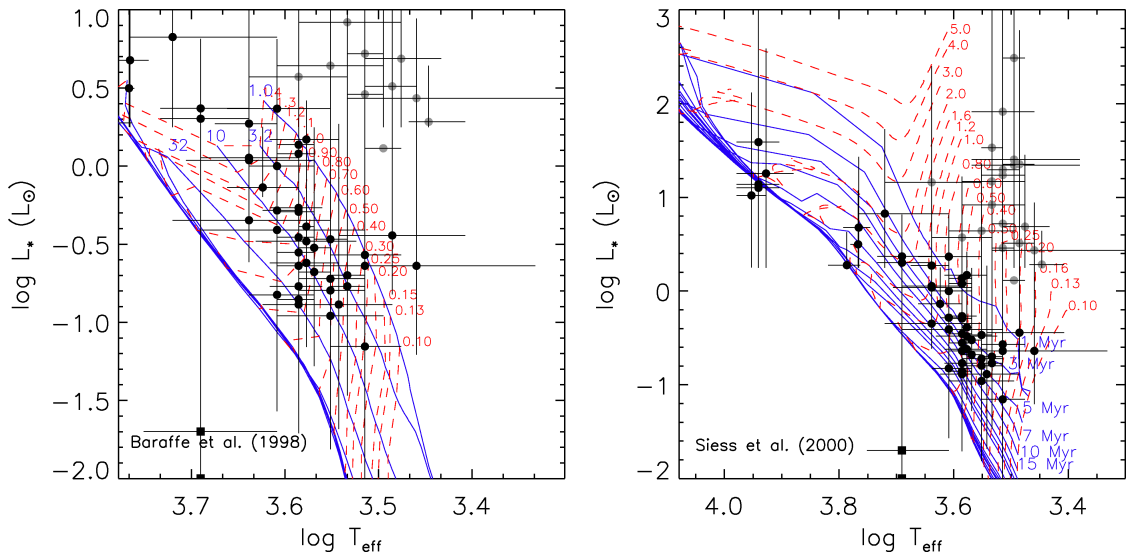}
\end{center}
\caption{\label{fhrd}H-R diagrams of the objects in this sample. Overlaid are the isochrones (blue) and mass tracks (red) from \citet{BA98} (left) and \citet{SI00} (right), with their corresponding age (in Myr) and mass (in M$_\odot$) indicated. Black circles correspond to objects in the Serpens Cloud, while gray circles are background sources (see text for details). Note the different scales in the two plots. }
\end{figure*}

\noindent
The H-R diagram of the sample is shown in Figure \ref{fhrd}. The stars are compared with evolutionary tracks from \citet{BA98} and \citet{SI00}. 20 objects are outside the range of the isochrones in the HR diagram (gray circles in Figure \ref{fhrd}). These objects are too luminous to be at a distance of 259 pc even in the extreme case of having vanishing extinction. As discussed above, their SEDs indeed show very little infrared excess, only at the longest wavelengths, and therefore they likely do not belong to the cloud. Obviously for these objects, no ages or masses could be determined. 

16 of these outliers have $A_V$Obs $>> A_V$Cloud and therefore appear in the shaded area in Figure \ref{extinction}. Thus, there is a 92\% correspondence of both methods to identify background sources. The 20 background stars are removed from the sample, leaving the remainder as a purer YSO sample in Serpens.

These remaining objects, shown with black circles in Figure \ref{fhrd}, are found to be consistently along the 1 to 10 Myr isochrones and between the 0.2 and 1.2 M$_\odot$ mass tracks in both diagrams, indicating a possibly coeval population of young stellar objects dominated by very low-mass stars. Early type ($T_{{\rm eff}} > 6000$K) stars fall outside the \citet{BA98} tracks, but are in the range of the \citet{SI00} models. In these cases, ages and masses could only be derived from the latter.

\begin{figure}[h!]
\begin{center}
\includegraphics[width=0.7\textwidth]{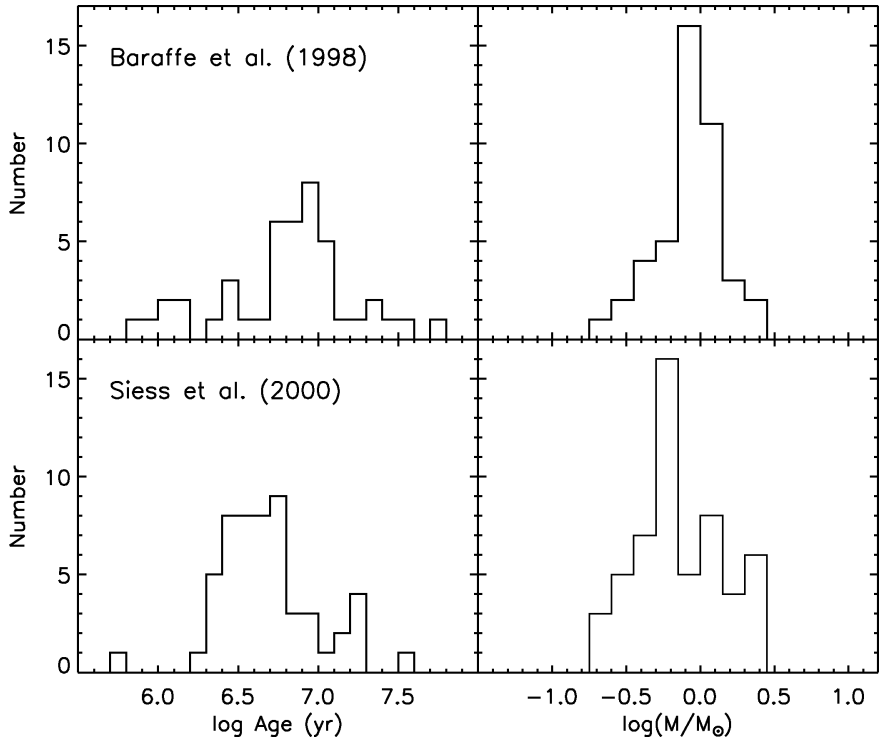}
\end{center}
\caption{\label{fage}Histograms of masses and ages for the sample of YSOs in Serpens, derived from the models of \citet{BA98} (top) and \citet{SI00} (bottom). }
\end{figure}

Individual ages and masses of the YSOs in Serpens were derived from the models and are shown in Table \ref{t_hrd} and Figure \ref{fage}. The \citet{BA98} tracks tend to produce larger masses and ages than those found with the \citet{SI00} tracks for the same position in the diagram. We also find that, for the different models, the differences in ages are within the luminosity errors, while the differences in masses are slightly larger than those due to the errors. In spite of the differences between models, we find similar mass and age distributions in both cases. Both sets of tracks imply a population of YSOs concentrated between 1 and 15 Myr with a few sources older than 15 Myr, and strongly peaked at 2 -- 6 Myr. The inferred median age is 7.5 Myr with the \citet{BA98} and 4.7 Myr with the \citet{SI00} tracks. Individual masses range from 0.2 to 2.2 M$_\odot$, with median values of 0.8 and 0.6 M$_\odot$ for the \citet{BA98} and \citet{SI00} models, respectively. The objects more massive than 2.0 M$_\odot$ (\#62, 70, 98, 108, 120, 139, 141) do not lie in the more clustered regions of the cloud, and are also separate from each other. The main uncertainty is caused by the uncertainty in extinction: higher (lower) extinction values by 2 mag can shift the \citet{SI00} median age by a factor of two to younger (older) ages. It is also important to stress that this is not a complete IMF study -- we observe a bias towards a larger mean age due to the bias in the selection of the sources (i.e. lack of very young, deeply embedded, Class 0 objects). Nevertheless, given that some of the stellar ages are clearly larger than the theoretical median age of a molecular cloud \citep{BA07}, we cannot rule out from the current age distribution that several star-forming events have taken place in the observed Serpens cloud. This is also consistent with the studied region in Serpens being larger than other coeval star-forming clusters, where less disperse age distributions have been found, e.g. IC 348 \citep{MU07} or Chamaeleon I \citep{LU07}.

Some basic tests for environmental trends were performed using our current dataset consisting of 78 stars. No strong correlations between the main parameters derived from the stellar spectra (ages and masses) and their positions in the cloud were found. This analysis will be revisited once the entire sample of 150 YSOs will be available.

Recent analysis suggests that the distance to Serpens may be as low as 193 $\pm$ 13 pc (Jens Knude, private communication), rather than the distance of 259 pc assumed here. The smaller distance would imply a notable increase in individual ages due to the decrease in luminosities, with median ages of 15.9 and 10.1 Myr with the \citet{BA98} and \citet{SI00} tracks, respectively. Median masses are not so strongly affected due to the mass tracks being almost vertical in the low-mass regime, yielding median masses of 0.8 and 0.6 M$_\odot$ according to the \citet{BA98} and \citet{SI00} tracks, respectively.

\section{Accretion Based on H$\alpha$ Emission}
\label{sacret}

\noindent
A relation between the H$\alpha$ line-to-continuum ratio and evolution of its emitting source has been widely explored (\citealt{BE89}, \citealt{HA98}, \citealt{MU03}, \citealt{WB03}, \citealt{SP04}, \citealt{NA04}). It is argued that the strength of the H$\alpha$ line decreases with evolutionary stage. A similar decline is seen in the strength of the prominent Ca II H and K lines (3968 and 3934 \AA{}, respectively) and in the surface rotational velocity, re-enforcing the motivation for a chromospheric origin for main-sequence H$\alpha$ emission. However, active T Tauri stars have H$\alpha$ fluxes much larger than those predicted by chromospheric models \citep{SP04}, meaning that for these objects H$\alpha$ is unlikely to originate primarily in a chromosphere. Mass accretion from a circumstellar disk is thought to be responsible for the excess emission of classical T Tauri stars (CTTS) \citep{LP74}. The inner disk is disrupted by the stellar magnetic field, resulting in magnetospheric accretion, where disk material is channelled along the magnetic field lines into the star \citep{KO91}. Prominent emission lines originate in the free-falling gas \citep{CH92}. H$\alpha$ is one of the strongest emission lines in CTTSs, and has been extensively studied as an indicator of accretion (e.g. \citealt{WB03}, \citealt{MU03}, \citealt{NA04}).

We use the H$\alpha$ line, covered by our optical spectra and present in emission in 35 objects of this sample, to obtain an indication of accretion activities according to two different methods from the literature: the H$\alpha$ equivalent width (EW) and its full width at 10\% of the peak intensity. The non-detection of H$\alpha$ is not uncommon for pre-main-sequence stars. The sources could either be more evolved WTTSs (see \S~\ref{sew}), or the non-detection could also be explained by the known temporal variability of T Tauri stars \citep{SP04}.

\subsection{H$\alpha$ Equivalent Width}
\label{sew}

\noindent
According to \citet{WB03}, there is no unique EW[H$\alpha$] value to distinguish between accreting CTTSs and weak-line T Tauri stars (WTTS). For instance, H$\alpha$ emission from equally saturated chromospheres will appear much more prominently in a late-M star than in an early-K star due to the former's substantially diminished photospheric continuum near 6500 \AA{}.

Taking into account this spectral type dependence, \citet{WB03} proposed that a T Tauri star is classical if EW[H$\alpha$] $\geq$ 3 \AA\ for K0--K5 stars, EW[H$\alpha$] $\geq$ 10 \AA\ for K7--M2.5 stars, EW[H$\alpha$] $\geq$ 20 \AA\ for M3--M5.5 stars, and EW[H$\alpha$] $\geq$ 40 \AA\ for M6--M7.5 stars. These values were determined empirically from the maximum values of EW[H$\alpha$] for non-veiled T Tauri stars within each spectral type, from high-resolution spectra. Stars with values of EW[H$\alpha$] below these levels are not necessarily WTTSs, for which confirmation depends upon the Li abundance or other youth indicators, such as variability.

We have determined the equivalent width of H$\alpha$ with the usual fitting of Gaussian profiles to the lines and the results are given in Table \ref{t_alpha}. The H$\alpha$ feature lies near the 6567 \AA\ TiO band head, a strong feature in mid- and late-M stars. In low resolution spectra, the edge of this band becomes blended with the H$\alpha$ emission feature, leading to an underestimate of the continuum on the redward side and thus an overestimate of the EW[H$\alpha$]. Also, self-absorption is missed in these unresolved data. For these reasons, our EW[H$\alpha$] values are likely somewhat overestimated. Nonetheless, we identify 35 stars with large EW[H$\alpha$] to be classified as CTTSs. The remaining four stars present marginal detections.

\subsection{Full Width of H$\alpha$ at 10\%}
\label{s10}

\noindent
Due to the uncertainties in the determination of EW[H$\alpha$] in low resolution spectra, \citet{WB03} proposed the full width of H$\alpha$ at 10\% of the peak intensity (H$\alpha$[10\%]) as an alternative to distinguish CTTSs from WTTSs. They found that stars with H$\alpha$[10\%] greater than 270 km s$^{-1}$ are CTTSs, independent of spectral types.

\begin{figure}[h!]
\begin{center}
\includegraphics[width=0.7\textwidth]{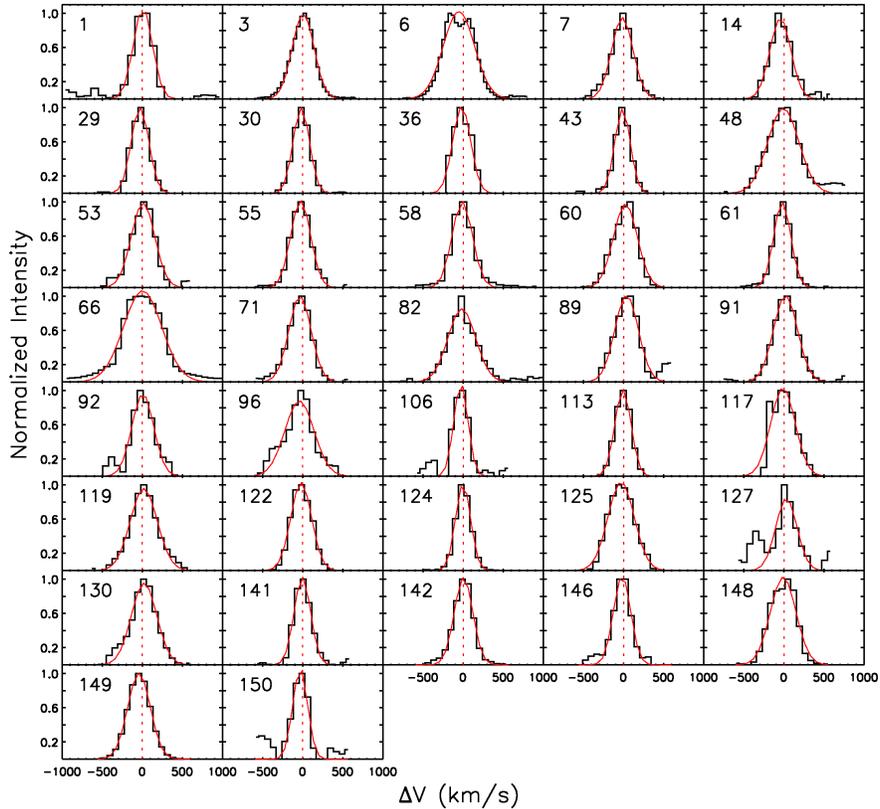}
\end{center}
\caption{\label{ha_ac} Continuum-subtracted profiles of H$\alpha$ emission lines. The solid red lines are the Gaussian fit to each profile, dashed red lines mark the center of the Gaussian. }
\end{figure}

We fitted Gaussian profiles to the H$\alpha$ lines (shown in Figure \ref{ha_ac}) in order to determine the H$\alpha$[10\%]. The errors were calculated by propagating the error on the spectral resolution and the error of the gaussian fit to the H$\alpha$ line. The measured FWHM of each profile was deconvolved assuming a gaussian instrumental profile. Because of the low resolution of our spectra (183.6 km s$^{-1}$ for {\it WYFFOS}, 126 km s$^{-1}$ for {\it DOLORES}, and 100 km s$^{-1}$ for {\it CAFOS}), very narrow profiles (e.g. objects 106 and 150) are unresolved. 

In spite of the low resolution, the H$\alpha$ profiles found are broad enough in most cases to allow the extraction of velocity information, although with low accuracy. These results are shown in Table \ref{t_alpha}. According to this criterion, 33 objects are CTTSs, while the remaining four stars are consistent with being CTTSs, given their uncertainties, showing a good agreement between both methods described.

\noindent
Studying accretion properties in low mass objects, \citet{NA04} found that H$\alpha$ can be used not only as a qualitative indicator of the accreting nature of low mass objects, but also to obtain a quantitative estimate of the mass accretion rate $\dot{M}_{ac}$. They found an empirical relationship between the mass accretion rate, $\dot{M}_{ac}$ (from H$\alpha$ profile model fittings or veiling measurements) and H$\alpha$[10\%] from high resolution spectra (Figure 3 in their paper):

\begin{equation}
\label{eac}
{\rm log}\dot{M}_{{\rm ac}} = -12.89(\pm0.3) + 9.7(\pm0.7) \times 10^{-3} \rm{H}\alpha [10\%]
\end{equation}

\noindent
where H$\alpha$[10\%] is in km s$^{-1}$ and $\dot{M}_{ac}$ is in M$_\odot$yr$^{-1}$. We use Eq. \ref{eac} to derive $\dot{M}_{{\rm ac}}$ for the objects showing H$\alpha$ in emission (see Table \ref{t_alpha}). 

\begin{figure}[h!]
\begin{center}
\includegraphics[width=0.6\textwidth]{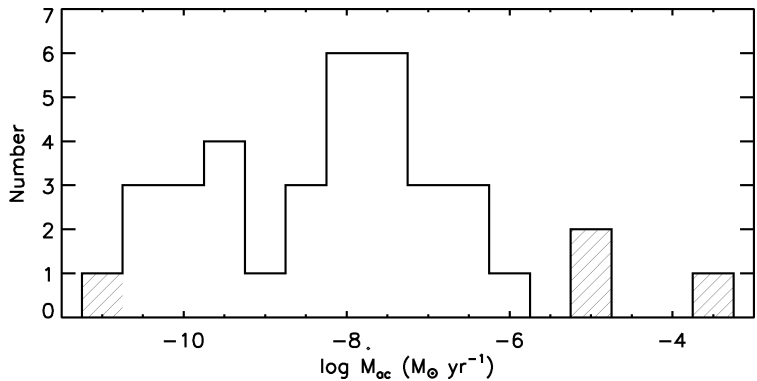}
\end{center}
\caption{\label{mac} Histogram of mass accretion rates derived from Eq. (\ref{eac}) for the sample of sources in Serpens with H$\alpha$ in emission. The shaded areas show objects out of the range for which the relationship between H$\alpha$[10\%] and $\dot{M}_{{\rm ac}}$ was calibrated by \citet{NA04}.  }
\end{figure}

The $\dot{M}_{{\rm ac}}$ values derived here (presented in Figure \ref{mac}) are largely consistent with the range typical for T Tauri stars. Exception are objects 109, 50, 7 and 68 (H$\alpha$[10\%] = 175, 762, 805 and 966 km s$^{-1}$, respectively). These H$\alpha$[10\%] values are out of the range over which relationship (\ref{eac}) was calibrated, which may result in erroneous accretion rates (shaded areas in Figure \ref{mac}). The distribution of mass accretion rates versus ages shows a large scatter, although for the 7 strongest accreters (log$\dot{M}_{ac} \gtrsim -7$) a weak trend is seen, in the sense of a decreasing accretion rate as a function of stellar age.

\section{Conclusions}
\label{scon}

\noindent
We have presented a spectroscopic survey at optical wavelengths designed to determine spectral types and confirm the pre-main sequence nature of a sample of young stellar object candidates in the Serpens Molecular Cloud, selected on the basis of the ``c2d'' Spitzer Legacy Program \citep{HA07}. This sample will subsequently be used to study disk evolution, making use of Spitzer infrared photometry and spectra, obtained by our group.

\begin{itemize}

\item We have determined spectral types for 78 stars in the cloud, finding mostly K and M stars (88\%), a late-type young stellar population. However, the high optical extinction toward this region makes it impossible to detect about half of the original sample of 150 stars with optical spectroscopy. Near-IR spectroscopy will be used to classify the remaining objects in the sample \citep{OL09}.

\item Comparison of extinction values determined from our optical spectra with those calculated from the contribution of the cloud at each point (averaged over a 5$'$ radius) is a powerful method for identifying candidate background sources with infrared excess from Spitzer selected samples of YSO candidates. The effectiveness of this method has been confirmed for two objects by follow-up IRS spectra. 

\item Effective temperatures and luminosities were derived and the objects were placed in a H-R diagram, compared with theoretical isochrones and mass tracks from models of \citet{BA98} and \citet{SI00}. 20 objects are too luminous to be at the distance of Serpens. These objects have a nearly one-to-one correspondence with those identified by the extinction method and are therefore very likely background objects. They amount to 25\% of the detected sample, a relatively large amount likely due to the position of the Serpens Molecular Cloud at very low Galactic latitude. Two of these background sources were confirmed to be post-AGB stars through IRS spectra.

\item The theoretical models by \citet{BA98} and \citet{SI00} imply a population of YSOs concentrated in the age range between 1 and 15 Myr, strongly peaked at 2 -- 6 Myr. The median age was found to be 7.5 and 4.7 Myr with the \citet{BA98} and \citet{SI00} tracks, respectively. Individual masses range from 0.2 to 2.2 M$_\odot$, with median values of 0.8 and 0.6 M$_\odot$ for the \citet{BA98} and \citet{SI00} models, respectively. 

\item The optical spectra cover the H$\alpha$ line, an important indicator of accretion. We explored this relationship through two different approaches: qualitatively, the accreting nature of low-mass objects has been determined by either the equivalent width of H$\alpha$, or its full width at 10\% of H$\alpha$ peak intensity. This confirms 37 objects (or 55\% of the YSO sample) to be actively accreting objects classified as new classical T Tauri stars in Serpens. The quantitative estimate of $\dot{M}_{{\rm ac}}$ based on the full width of H$\alpha$ at 10\% of the peak intensity yields a median of $\sim10^{-8}$ M$_\odot$yr$^{-1}$, with a broad distribution of values. 

\end{itemize}

\acknowledgments
Support for this work, part of the Spitzer Space Telescope Legacy Science Program, was provided by NASA through Contract Numbers 1256316, 1224608 and 1230780 issued by the Jet Propulsion Laboratory, California Institute of Technology under NASA contract 1407 and by the Spanish Grant AYA 2005-0954. Astrochemistry at Leiden is supported by a Nederlandse Organisatie voor Wetenschappelijk Onderzoek (NWO) Spinoza and Nederlandse Onderzoekschool voor de Astronomie (NOVA) grants, and by the European Research Training Network ``The Origin of Planetary Systems'' (PLANETS, contract number HPRN-CT-2002-00308). Support to KMP was provided by NASA through Hubble Fellowship grant \#01201.01, awarded by the Space Telescope Science Institute, which is operated by the Association of Universities for Research in Astronomy, Inc., for NASA, under contract NAS 5-26555. 
The authors would like to thank N.J. Evans II, L. Spezzi, J. Alcal\'a, G. Herczeg, J. Falc\'on-Barroso, D. Barrado y Navascues, and A. Bayo for helpful comments and suggestions on earlier versions of the manuscript.

\clearpage
\thispagestyle{empty}
\setlength{\voffset}{30mm}
{\rotate
\begin{table}
\caption{Observation log}      
\begin{center}
\begin{tabular}{cccccp{7.cm}}
\hline
\hline
\tabletypesize{\footnotesize}    
Date & Telescope + & Wavelength  & Resolution  & Spectral  & \    \    \ Object ID (as in Table \ref{t_id}) \\
 & Instrument & Blaze (\AA) & (\AA) & Coverage (\AA) & \\
\hline
2006 May 4          &     WHT+WYFFOS   & 7000 & 4.0 & 3000 & 16, 34, 44, 47, 48, 87, 89, 91, 93, 96, 97, 113, 119, 123, 125, 128, 130, 145, 148  \\
2006 June 29, 30    &    TNG+DOLORES   & 7000 & 2.8 & 1600 & 2, 3, 4, 6, 35, 52, 70, 76, 85, 86, 108, 118, 120, 122, 131, 139, 146, 150  \\
2007 June 9, 10, 11 &     WHT+WYFFOS   & 7000 & 4.0 & 3000 & 2, 4, 5, 7, 14, 16, 18, 22, 23, 29, 30, 35, 36, 40, 43, 52, 53, 55, 60, 61, 62, 66, 70, 71, 76, 77, 81, 82, 84, 85, 86, 87, 88, 92, 93, 96, 98, 99, 106, 110, 112, 113, 114, 117, 118, 119, 120, 121, 122, 124, 125, 127, 130, 131, 138, 139, 141, 142, 146, 148, 149, 150  \\
2008 June 29, 30    & Calar Alto+CAFOS & 7750 & 2.2 & 3500 & 1, 41, 58, 70, 82, 93, 99, 115, 120, 130, 139, 141, 149, 150 \\
\hline
\end{tabular}
\end{center}
\label{t_obs}
\end{table}
}
\clearpage
\setlength{\voffset}{0mm}

\begin{deluxetable}{l l c c c c c c }
\tabletypesize{\footnotesize}
\tablecolumns{8}
\tablewidth{0pt}
\tablecaption{Stellar parameters for the YSO candidates in Serpens \label{t_id}} 
\tablehead{\colhead{\#}                  & 
           \colhead{ID\tablenotemark{a}}               & 
           \colhead{c2d ID}               &  
           \colhead{$\alpha_{2\mu{\rm m}-24\mu{\rm m}}$\tablenotemark{b}}  &  
           \colhead{Spectral Type}        &
           \colhead{$A_V$Obs\tablenotemark{c}}         &  
           \colhead{$A_V$Cloud\tablenotemark{d}}       & 
           \colhead{$A_V$Final}           \\
           \colhead{}                     & 
           \colhead{}                     & 
           \colhead{SSTc2dJ}              & 
           \colhead{}                     &
           \colhead{($\pm$subtype)}       & 
           \colhead{(mag)}                & 
           \colhead{(mag)}                &
           \colhead{(mag)}                }
\startdata
     1 &   1      &  18275383-0002335    &   -1.26  &     K2$\pm$3     &     1.0   &    2.98   &    9.0  \\
     2 &   2      &  18280503+0006591    &   -2.29  &     M4$\pm$1     &     6.5   &    1.98   &    2.0  \\ 
     3 &   3      &  18280845-0001064    &   -0.70  &     M0$\pm$1     &     0.0   &    2.85   &    2.0  \\ 
     4 &   4      &  18281100-0001395    &   -2.13  &     M5$\pm$1     &     8.0   &    2.85   &    6.0  \\ 
     5 &          &  18281315+0003128    &   -1.80  &     M3$\pm$3     &    14.0   &    4.83   &    9.0  \\ 
     6 &   5      &  18281350-0002491    &   -1.08  &     K5$\pm$2     &     2.0   &    5.33   &    3.0  \\ 
     7 &   6      &  18281501-0002588    &   -0.05  &     M0$\pm$1     &     2.0   &    5.33   &    7.0  \\ 
    14 &   13     &  18282143+0010411    &   -1.44  &     M2$\pm$2     &     3.5   &    2.66   &    3.0  \\ 
    16 &   15     &  18282432+0034545    &   -2.01  &     M0$\pm$1     &     8.0   &    3.41   &    6.0  \\ 
    18 &   16     &  18282738-0011499    &   -1.98  &     M5$\pm$4     &    10.0   &    4.78   &   11.0  \\ 
    22 &          &  18283000+0020147    &   -1.60  &     M4$\pm$3     &    14.0   &    4.74   &   12.0  \\ 
    23 &          &  18283736+0019276    &   -2.19  &   M7.5$\pm$1     &     0.0   &    7.59   &   11.0  \\ 
    29 &   25     &  18284481+0048085    &   -1.01  &     M2$\pm$3     &     3.0   &    4.29   &    6.0  \\  
    30 &   27     &  18284497+0045239    &   -1.35  &     M1$\pm$1     &     2.5   &    7.55   &    2.0  \\ 
    34 &          &  18284828-0005300    &   -2.43  &   M5.5$\pm$1.5   &    11.0   &    4.92   &   10.0  \\ 
35$^e$ &          &  18284938-0006046    &   -2.01  &     M5$\pm$1     &    12.0   &    5.72   &   13.0  \\ 
    36 &   32     &  18285020+0009497    &   -0.24  &     K5$\pm$4     &     5.0   &    5.40   &   10.0  \\ 
    40 &   36     &  18285249+0020260    &   -0.15  &     M7$\pm$4     &     8.5   &   10.00   &   12.0  \\ 
    41 &   37     &  18285276+0028466    &    0.06  &     K2$\pm$3     &     0.0   &   19.20   &   10.0  \\
    43 &   39     &  18285395+0045530    &   -1.16  &   M0.5$\pm$3     &     3.5   &    8.49   &    7.0  \\ 
    44 &   40     &  18285404+0029299    &    1.35  &     M6$\pm$5     &     0.0   &   15.60   &         \\ 
    47 &   43     &  18285489+0018326    &    0.88  &     M5$\pm$4     &    13.5   &   14.20   &         \\ 
    48 &   44     &  18285529+0020522    &   -0.29  &   M5.5$\pm$3     &     3.0   &    6.86   &   14.0  \\ 
    52 &   48     &  18285808+0017244    &   -2.11  &     G3$\pm$5     &     3.5   &    9.87   &    6.0  \\ 
    53 &   49     &  18285860+0048594    &   -1.06  &   M2.5$\pm$3     &     2.5   &    3.37   &    6.0  \\  
    55 &   51     &  18290025+0016580    &   -1.01  &     K2$\pm$3     &     4.5   &    7.88   &   12.0  \\
    58 &   54     &  18290088+0029315    &   -0.42  &     K7$\pm$2     &     3.5   &   19.00   &    5.0  \\
    60 &   56     &  18290122+0029330    &   -0.39  &   M0.5$\pm$0.5   &     3.0   &   19.00   &    7.0  \\ 
    61 &   58     &  18290175+0029465    &   -0.84  &     M0$\pm$1     &     2.0   &   19.00   &    5.0  \\ 
    62 &   59     &  18290184+0029546    &   -0.77  &     K0$\pm$7     &     4.0   &   19.00   &    8.0  \\ 
    66 &   62     &  18290393+0020217    &   -0.82  &     K5$\pm$2     &     1.5   &    3.53   &    7.0  \\ 
    70 &   65     &  18290575+0022325    &   -2.12  &     A3$\pm$3     &     2.5   &    4.17   &    5.0  \\ 
    71 &   66     &  18290615+0019444    &   -1.52  &     M3$\pm$1     &     2.0   &    4.06   &    6.0  \\ 
    76 &   72     &  18290775+0054037    &   -1.06  &     M1$\pm$1     &     2.0   &    2.32   &    6.0  \\ 
    77 &   73     &  18290808-0007371    &   -2.22  &     M4$\pm$1     &     8.5   &    3.10   &    8.0  \\ 
    81 &   77     &  18290980+0034459    &   -0.97  &     M5$\pm$4     &     1.0   &   11.00   &   15.0  \\ 
    82 &   78     &  18291148+0020387    &   -1.34  &     M0$\pm$2     &     2.0   &    3.30   &    6.0  \\ 
    84 &   81     &  18291407+0002589    &   -2.19  &     M3$\pm$3     &     9.0   &    4.74   &    9.0  \\ 
85$^e$ &          &  18291477-0004237    &   -2.29  &     M3$\pm$1     &     8.0   &    2.71   &    9.0  \\ 
    86 &   83     &  18291508+0052124    &   -1.89  &   M5.5$\pm$1.5   &     8.0   &    2.51   &    8.0  \\ 
    87 &   84     &  18291513+0039378    &   -1.50  &     M4$\pm$3     &     2.5   &    7.95   &    7.0  \\ 
    88 &   85     &  18291539-0012519    &   -1.72  &   M0.5$\pm$2     &     7.0   &    2.51   &    7.0  \\ 
    89 &   86     &  18291557+0039119    &   -0.56  &     K5$\pm$5     &     2.0   &   10.20   &   12.0  \\ 
    91 &   88     &  18291617+0018227    &    0.45  &     K7$\pm$3     &     1.0   &    4.97   &   18.0  \\ 
    92 &   89     &  18291969+0018031    &   -1.23  &     M0$\pm$1     &     2.5   &    3.51   &    7.0  \\ 
    93 &   90     &  18292001+0024497    &   -2.04  &     M2$\pm$3     &     8.5   &    3.17   &    9.0  \\ 
    96 &   94     &  18292184+0019386    &   -0.99  &     M1$\pm$1     &     2.0   &    3.16   &    7.0  \\ 
    97 &          &  18292250+0034118    &   -1.65  &     M2$\pm$2     &     3.0   &   10.10   &   10.1  \\ 
    98 &          &  18292253+0034176    &   -2.48  &     A3$\pm$5     &     5.5   &   10.10   &    5.0  \\ 
    99 &   95     &  18292616+0020518    &   -2.26  &     M4$\pm$2     &    13.0   &    3.34   &    8.0  \\ 
   106 &   102    &  18292927+0018000    &   -1.24  &     M3$\pm$1     &     2.5   &    3.48   &    7.5  \\ 
   108 &   105    &  18293254-0013233    &   -2.58  &     K5$\pm$5     &     0.0   &    2.51   &    1.0  \\ 
   110 &   107    &  18293319+0012122    &   -1.81  &     M6$\pm$2     &     8.5   &    3.90   &    5.0  \\ 
   112 &   109    &  18293381+0053118    &   -2.06  &     M7$\pm$4     &     1.5   &    4.59   &    8.0  \\ 
   113 &   110    &  18293561+0035038    &   -1.53  &     K7$\pm$1     &     2.0   &   12.60   &    4.0  \\ 
   114 &   111    &  18293619+0042167    &   -0.92  &     F9$\pm$5     &     7.0   &    6.71   &   11.0  \\ 
   115 &   114    &  18293672+0047579    &   -1.08  &   M0.5$\pm$2     &     6.5   &    5.03   &    8.0  \\
   117 &   118    &  18294020+0015131    &    0.87  &     K2$\pm$5     &     0.5   &    3.40   &    2.0  \\
   118 &          &  18294067-0007033    &   -2.53  &     M4$\pm$1.5   &     8.0   &    1.41   &    8.0  \\ 
   119 &   119    &  18294121+0049020    &   -1.39  &     K7$\pm$2     &     4.0   &    5.20   &    5.5  \\ 
   120 &   123    &  18294168+0044270    &   -1.42  &     A2$\pm$2     &     5.0   &    6.13   &    8.5  \\ 
   121 &          &  18294301-0016083    &   -2.18  &     K7$\pm$5     &     5.5   &    3.33   &    7.0  \\
   122 &   126    &  18294410+0033561    &   -1.68  &     M0$\pm$1.5   &     1.5   &    7.35   &    5.0  \\ 
   123 &   129    &  18294503+0035266    &   -1.26  &     M0$\pm$1     &     1.5   &    7.02   &   10.0  \\ 
   124 &   132    &  18294725+0039556    &   -1.32  &     M0$\pm$1.5   &     2.0   &   10.40   &    5.5  \\ 
   125 &   133    &  18294726+0032230    &   -1.31  &     M0$\pm$1     &     3.0   &    6.59   &    7.0  \\ 
   127 &   143    &  18295001+0051015    &   -1.68  &     M2$\pm$1     &     4.0   &    4.18   &    6.5  \\ 
   130 &   145    &  18295041+0043437    &   -1.32  &     K6$\pm$1     &     4.5   &    8.69   &    8.5  \\ 
   131 &   148    &  18295130+0027479    &   -2.42  &     A3$\pm$3     &     2.0   &    3.34   &    5.5  \\ 
   138 &   161    &  18295322+0033129    &   -1.99  &     M0$\pm$3     &    10.0   &    6.86   &   11.0  \\ 
   139 &   165    &  18295422+0045076    &   -2.35  &     A4$\pm$4     &     4.0   &    6.95   &    6.5  \\ 
   141 &   169    &  18295531+0049393    &   -0.74  &     A3$\pm$3     &     3.5   &    3.02   &    9.0  \\ 
   142 &   172    &  18295592+0040150    &   -0.85  &     M4$\pm$2     &     3.5   &    5.71   &    3.5  \\ 
   145 &   178    &  18295714+0033185    &   -0.82  &   G2.5$\pm$2.5   &     3.5   &    5.19   &   10.0  \\ 
   146 &   182    &  18295772+0114057    &    0.44  &     M4$\pm$2     &     4.0   &   14.05   &    3.0  \\ 
   148 &   206    &  18300178+0032162    &   -1.25  &     K7$\pm$1     &     2.0   &    5.00   &    7.0  \\ 
   149 &   210    &  18300350+0023450    &   -1.22  &     M0$\pm$1     &     0.0   &    2.25   &    3.5  \\ 
   150 &   222    &  18300861+0058466    &   -0.70  &     K5$\pm$3     &     0.0   &    2.79   &    5.0  \\
\enddata
\tablenotetext{a}{From \citet{HA07}}
\tablenotetext{b}{Obtained from a linear fit to the logarithm of the fluxes between the 2MASS $K$ (2$\mu$m) and the MIPS1 (24$\mu$m) bands \citep{HA07}}
\tablenotetext{c}{Extinction found in this work}
\tablenotetext{d}{Extinction found by ``c2d'' over 5$'$ region}
\tablenotetext{e}{post-AGB stars confirmed with Spitzer IRS spectra}
\end{deluxetable}

\begin{deluxetable}{l c c c c c c }
\tabletypesize{\footnotesize}
\tablecolumns{7}
\tablewidth{0pt}
\tablecaption{Stellar parameters for the YSO candidates in Serpens \label{t_hrd}} 
\tablehead{\colhead{\#}              & 
           \colhead{T$_{{\rm eff}}$}  & 
           \colhead{L$_{*}$}          & 
           \colhead{Age$_B$}          & 
           \colhead{Mass$_B$}         &
           \colhead{Age$_S$}          & 
           \colhead{Mass$_S$}         \\
           \colhead{}                 &
           \colhead{(K)}              & 
           \colhead{(L$_\odot$)}      & 
           \colhead{(Myr)}            & 
           \colhead{(M$_\odot$)}      & 
           \colhead{(Myr)}            & 
           \colhead{(M$_\odot$)}      }
\startdata
1                           &  4900 $^{+510}_{-550}$     &      2.01$^{+2.39}_{-1.13}$   &  11.60$^{+13.91}_{-10.01}$  & 1.56$^{+0.63}_{-0.48}$ &  6.10$^{+11.00}_{-4.15}$    & 1.57$^{+0.43}_{-0.44}$  \\ 
2\tablenotemark{\dagger}    &  3270 $\pm$145             &      5.24$^{+6.26}_{-2.95}$   &                             &                        &                             &                         \\ 
3                           &  3850 $^{+210}_{-145}$     &      0.51$^{+0.61}_{-0.29}$   &   5.42$^{+10.98}_{-3.67}$   & 0.92$^{+0.13}_{-0.16}$ &  2.76$^{+5.58}_{-1.98}$     & 0.57$^{+0.22}_{-0.11}$  \\ 
4\tablenotemark{\dagger}    &  3125 $^{+145}_{-135}$     &      1.30$^{+1.55}_{-0.73}$   &                             &                        &                             &                         \\ 
5\tablenotemark{\dagger}    &  3415 $^{+435}_{-425}$     &     14.78$^{+17.64}_{-8.31}$  &                             &                        &                             &                         \\ 
6                           &  4350 $^{+380}_{-290}$     &      1.13$^{+1.35}_{-0.64}$   &   5.98$^{+13.19}_{-4.28}$   & 1.38$^{+0.18}_{-0.34}$ &  3.41$^{+8.77}_{-2.10}$     & 1.13$^{+0.19}_{-0.37}$  \\ 
7                           &  3850 $^{+210}_{-145}$     &      0.23$^{+0.27}_{-0.13}$   &  16.25$^{+35.98}_{-10.56}$  & 0.81$^{+0.11}_{-0.15}$ &  8.26$^{+21.22}_{-5.45}$    & 0.59$^{+0.19}_{-0.12}$  \\ 
14                          &  3560 $\pm$290             &      0.19$^{+0.22}_{-0.10}$   &   6.87$^{+14.79}_{- 4.99}$  & 0.52$\pm$0.25          &  4.37$^{+7.90}_{-2.00}$     & 0.37$^{+0.22}_{-0.12}$  \\ 
16                          &  3850 $^{+210}_{-145}$     &      1.20$^{+1.44}_{-0.68}$   &   1.56$^{+3.64}_{- 1.36}$   & 1.03$^{+0.16}_{-0.17}$ &  0.52$^{+2.19}_{-4.64}$     & 0.55$^{+0.20}_{-0.09}$  \\ 
18\tablenotemark{\dagger}   &  3125 $^{+580}_{-725}$     &     25.26$^{+30.15}_{-14.21}$ &                             &                        &                             &                         \\ 
22\tablenotemark{\dagger}   &  3270 $^{+435}_{-390}$     &     81.97$^{+97.83}_{-46.11}$ &                             &                        &                             &                         \\ 
23\tablenotemark{\dagger}   &  2795 $^{+140}_{-240}$     &      1.92$^{+2.29}_{-1.08}$   &                             &                        &                             &                         \\ 
29                          &  3560 $^{+500}_{-435}$     &      0.11$^{+0.14}_{-0.06}$   &  12.51$^{+62.38}_{-11.01}$  & 0.50$^{+0.16}_{-0.35}$ &  7.80$^{+46.64}_{-4.83}$    & 0.36$^{+0.29}_{-0.19}$  \\  
30                          &  3705 $\pm$145             &      0.30$^{+0.36}_{-0.17}$   &   6.57$^{+13.95}_{-4.33}$   & 0.71$^{+0.15}_{-0.16}$ &  3.41$^{+6.76}_{-1.63}$     & 0.46$^{+0.12}_{-0.09}$  \\ 
34\tablenotemark{\dagger}   &  3057 $^{+212}_{-177}$     &      3.24$^{+3.87}_{-1.82}$   &                             &                        &                             &                         \\ 
35\tablenotemark{a,\dagger} &  3125 $^{+145}_{-135}$     &  306.47$^{+365.79}_{-172.40}$ &                             &                        &                             &                         \\ 
36                          &  4350 $^{+730}_{-645}$     &      1.09$^{+1.30}_{-0.61}$   &   6.34$^{+19.20}_{-5.33}$   & 1.36$^{+0.20}_{-0.51}$ &  3.68$^{+14.43}_{-3.46}$    & 1.13$^{+0.03}_{-0.67}$  \\ 
40                          &  2880 $^{+535}_{-730}$     &      0.23$^{+0.27}_{-0.13}$   &                             &                        &  2.86$^{+2.86}_{-2.80}$     & 0.31$^{+0.31}_{-0.30}$  \\ 
41                          &  4900 $^{+510}_{-550}$     &      0.01$\pm$0.01            &                             &                        &                             &                         \\
43                          &  3777 $^{+500}_{-435}$     &      0.33$^{+0.39}_{-0.18}$   &   7.47$^{+19.56}_{-6.17}$   & 0.80$^{+0.11}_{-0.44}$ &  3.79$^{+14.75}_{-2.02}$    & 0.52$^{+0.39}_{-0.23}$  \\ 
44\tablenotemark{b}         &  2990 $^{+715}_{-840}$     &                               &                             &                        &                             &                         \\ 
47\tablenotemark{b}         &  3125 $^{+580}_{-725}$     &                               &                             &                        &                             &                         \\ 
48                          &  3057 $^{+430}_{-502}$     &      0.36$^{+0.42}_{-0.20}$   &                             &                        &  2.47$^{+2.47}_{-2.40}$     & 0.35$^{+0.35}_{-0.30}$  \\ 
52                          &  5830 $^{+370}_{-260}$     &      4.76$^{+5.68}_{-2.68}$   &   8.70$^{+15.63}_{-8.02}$   & 2.53$^{+0.64}_{-1.16}$ & 12.00$^{+9.69}_{-6.27}$     & 1.50$^{+0.46}_{-0.28}$  \\ 
53                          &  3487 $^{+537}_{-430}$     &      0.13$^{+0.16}_{-0.07}$   &   7.52$^{+49.36}_{-6.95}$   & 0.44$^{+0.27}_{-0.31}$ &  5.35$^{+34.68}_{-3.54}$    & 0.33$^{+0.35}_{-0.17}$  \\  
55                          &  4900 $^{+510}_{-550}$     &      2.34$^{+2.79}_{-1.31}$   &   9.45$^{+12.25}_{-8.31}$   & 1.60$^{+0.53}_{-0.45}$ &  4.98$^{+9.13}_{-3.47}$     & 1.66$^{+0.41}_{-0.53}$  \\
58                          &  4060 $^{+290}_{-355}$     &      1.00$^{+1.19}_{-0.56}$   &   3.15$^{+7.30}_{-2.20}$    & 1.18$^{+0.15}_{-0.34}$ &  2.39$^{+3.91}_{-2.33}$     & 0.76$^{+0.37}_{-0.30}$  \\
60                          &  3777 $\pm$72              &      0.41$^{+0.48}_{-0.23}$   &   5.73$^{+11.95}_{-3.81}$   & 0.82$^{+0.08}_{-0.09}$ &  2.89$^{+6.01}_{-1.71}$     & 0.52$^{+0.06}_{-0.05}$  \\ 
61                          &  3850 $^{+210}_{-145}$     &      1.37$^{+1.63}_{-0.77}$   &   1.24$^{+3.01}_{-1.12}$    & 1.05$^{+0.17}_{-0.16}$ &  2.07$\pm$0.41              & 0.66$\pm$0.09           \\ 
62                          &  5250 $^{+580}_{-1190}$    &      6.70$^{+7.99}_{-3.77}$   &   2.08$^{+4.18}_{-2.07}$    & 2.19$^{+1.17}_{-0.56}$ &  3.87$^{+4.94}_{-3.63}$     & 2.09$^{+0.49}_{-1.52}$  \\ 
66                          &  4350 $^{+380}_{-290}$     &      1.87$^{+2.23}_{-1.05}$   &   4.26$^{+4.79}_{-3.72}$    & 1.44$^{+0.25}_{-0.19}$ &  2.13$^{+3.49}_{-2.13}$     & 1.12$^{+0.42}_{-0.39}$  \\ 
70                          &  8720 $^{+800}_{-695}$     &     12.64$^{+15.09}_{-7.11}$  &                             &                        &  8.18$\pm$2.21              & 2.09$\pm$0.19           \\ 
71                          &  3415 $\pm$145             &      0.20$^{+0.24}_{-0.11}$   &   2.86$^{+5.90}_{-1.64}$    & 0.38$^{+0.14}_{-0.11}$ &  2.94$^{+3.96}_{-0.93}$     & 0.31$\pm$0.06           \\ 
76                          &  3705 $\pm$145             &      0.21$^{+0.25}_{-0.12}$   &   9.62$^{+21.62}_{-6.22}$   & 0.68$^{+0.11}_{-0.16}$ &  5.40$^{+11.80}_{-2.90}$    & 0.46$^{+0.13}_{-0.09}$  \\ 
77\tablenotemark{\dagger}   &  3270 $\pm$145             &      2.88$^{+3.44}_{-1.62}$   &                             &                        &                             &                         \\ 
81\tablenotemark{\dagger}   &  3125 $^{+580}_{-725}$     &     22.08$^{+26.35}_{-12.42}$ &                             &                        &                             &                         \\ 
82                          &  3850 $^{+355}_{-290}$     &      0.13$^{+0.15}_{-0.07}$   &  32.64$^{+56.97}_{-22.59}$  & 0.72$^{+0.13}_{-0.22}$ & 19.65$^{+57.48}_{-13.69}$   & 0.58$^{+0.09}_{-0.22}$  \\ 
84\tablenotemark{\dagger}   &  3415 $^{+435}_{-425}$     &      8.33$^{+9.94}_{-4.68}$   &                             &                        &                             &                         \\ 
85\tablenotemark{a,\dagger} &  3415 $\pm$145             &     33.83$^{+40.37}_{-19.03}$ &                             &                        &                             &                         \\ 
86\tablenotemark{\dagger}   &  3057 $^{+212}_{-177}$     &     22.52$^{+26.89}_{-12.67}$ &                             &                        &                             &                         \\ 
87                          &  3270 $^{+435}_{-390}$     &      0.23$^{+0.28}_{-0.13}$   &   1.36$^{+7.32}_{-1.36}$    & 0.28$^{+0.41}_{-0.21}$ &  2.57$^{+2.13}_{-2.57}$    & 0.25$^{+0.21}_{-0.08}$   \\ 
88                          &  3777 $^{+355}_{-290}$     &      1.48$^{+1.77}_{-0.83}$   &   0.74$^{+2.08}_{-0.64}$    & 0.97$^{+0.31}_{-0.14}$ &  1.76$^{+0.27}_{-0.27}$    & 0.66$^{+0.15}_{-0.15}$   \\ 
89                          &  4350 $^{+900}_{-790}$     &      0.45$^{+0.54}_{-0.25}$   &  21.77$^{+449.61}_{-19.55}$ & 1.00$^{+0.32}_{-0.42}$ & 13.98$^{+35.23}_{-13.10}$   & 1.00$^{+0.13}_{-0.61}$  \\ 
91                          &  4060 $^{+530}_{-500}$     &      2.33$^{+2.79}_{-1.31}$   &   0.82$^{+2.26}_{-0.80}$    & 1.28$^{+0.59}_{-0.24}$ &  2.53$\pm$0.20              & 1.13$\pm$0.37           \\ 
92                          &  3850 $^{+210}_{-145}$     &      0.28$^{+0.34}_{-0.16}$   &  10.82$^{+23.60}_{-6.86}$   & 0.85$^{+0.10}_{-0.15}$ &  5.82$^{+14.61}_{-3.42}$    & 0.58$^{+0.22}_{-0.12}$  \\ 
93\tablenotemark{\dagger}   &  3560 $^{+500}_{-435}$     &      4.39$^{+5.24}_{-2.47}$   &                             &                        &                             &                         \\ 
96                          &  3705 $\pm$145             &      0.21$^{+0.26}_{-0.12}$   &   9.46$^{+21.21}_{-6.18}$   & 0.68$^{+0.11}_{-0.16}$ &  5.23$^{+11.50}_{-2.76}$    & 0.46$^{+0.13}_{-0.09}$  \\ 
97                          &  3560 $\pm$290             &      0.16$^{+0.19}_{-0.09}$   &   8.01$^{+17.71}_{-5.82}$   & 0.51$^{+0.24}_{-0.25}$ &  4.94$^{+9.43}_{-2.37}$     & 0.37$^{+0.22}_{-0.12}$  \\ 
98                          &  8720 $^{+3180}_{-1140}$   &      5.44$^{+6.49}_{-3.06}$   &                             &                        &  3.79$^{+3.79}_{-3.79}$     & 2.73$^{+2.73}_{-2.73}$  \\ 
99\tablenotemark{\dagger}   &  3270 $^{+290}_{-280}$     &     17.27$^{+20.61}_{-9.72}$  &                             &                        &                             &                         \\ 
106                         &  3415 $\pm$145             &      0.17$^{+0.20}_{-0.09}$   &   3.48$^{+7.76}_{-1.86}$    & 0.37$^{+0.14}_{-0.10}$ &  3.63$^{+5.22}_{-1.31}$     & 0.30$\pm$0.06           \\ 
108\tablenotemark{\dagger}  &  4350 $^{+900}_{-790}$     &     14.48$^{+17.28}_{-8.15}$  &                             &                        &                             &                         \\ 
110\tablenotemark{\dagger}  &  2990 $\pm$280             &      4.87$^{+5.81}_{-2.74}$   &                             &                        &                             &                         \\ 
112\tablenotemark{\dagger}  &  2880 $^{+535}_{-880}$     &      2.72$^{+3.24}_{-1.53}$   &                             &                        &                             &                         \\
113                         &  4060 $^{+145}_{-210}$     &      0.52$^{+0.61}_{-0.29}$   &   8.64$^{+19.10}_{-5.88}$   & 1.06$^{+0.13}_{-0.22}$ &  4.78$^{+14.01}_{-2.65}$    & 0.79$^{+0.16}_{-0.22}$  \\ 
114                         &  6115 $^{+475}_{-315}$     &      1.89$^{+2.25}_{-1.06}$   &                             &                        & 19.37$\pm$3.64              & 1.30$\pm$0.09           \\ 
115                         &  3777 $^{+355}_{-290}$     &      0.24$^{+0.29}_{-0.14}$   &  10.28$^{+23.83}_{-7.21}$   & 0.77$^{+0.08}_{-0.32}$ &  5.70$^{+15.08}_{-3.25}$    & 0.52$^{+0.30}_{-0.17}$  \\
117                         &  4900 $^{+736}_{-840}$     &      0.02$^{+0.02}_{-0.01}$   &                             &                        &                             &                         \\
118\tablenotemark{\dagger}  &  3270 $^{+217}_{-212}$     &     19.67$^{+23.47}_{-11.06}$ &                             &                        &                             &                         \\ 
119                         &  4060 $^{+290}_{-355}$     &      0.15$^{+0.18}_{-0.08}$   &  51.52$^{+46.77}_{-51.52}$  & 0.73$^{+0.20}_{-0.07}$ & 36.46$^{+26.56}_{-36.40}$   & 0.70$^{+0.11}_{-0.25}$  \\ 
120                         &  8970 $^{+550}_{-510}$     &     10.48$^{+12.51}_{-5.89}$  &                             &                        &  7.22$^{+7.22}_{-7.17}$     & 2.19$^{+2.19}_{-2.14}$  \\ 
121                         &  4060 $^{+840}_{-790}$     &      0.39$^{+0.47}_{-0.22}$   &  12.68$^{+76.74}_{-12.21}$  & 0.99$^{+0.15}_{-0.69}$ &  7.38$^{+40.69}_{-5.73}$    & 0.81$^{+0.06}_{-0.55}$  \\
122                         &  3850 $^{+282}_{-217}$     &      0.54$^{+0.64}_{-0.30}$   &   5.06$^{+10.19}_{-3.44}$   & 0.93$^{+0.17}_{-0.25}$ &  2.68$^{+5.04}_{-2.07}$     & 0.57$^{+0.31}_{-0.15}$  \\ 
123                         &  3850 $^{+210}_{-145}$     &      0.35$^{+0.42}_{-0.20}$   &   8.54$^{+18.18}_{-5.66}$   & 0.88$^{+0.11}_{-0.15}$ &  4.34$^{+10.41}_{-2.43}$    & 0.58$^{+0.23}_{-0.11}$  \\ 
124                         &  3850 $^{+282}_{-217}$     &      0.17$^{+0.21}_{-0.10}$   &  23.46$^{+52.63}_{-15.55}$  & 0.76$^{+0.12}_{-0.17}$ & 12.65$^{+33.50}_{-8.73}$    & 0.59$^{+0.15}_{-0.18}$  \\ 
125                         &  3850 $^{+210}_{-145}$     &      0.28$^{+0.34}_{-0.16}$   &  10.98$^{+23.96}_{-6.97}$   & 0.85$^{+0.10}_{-0.15}$ &  5.89$^{+14.82}_{-3.47}$    & 0.58$^{+0.22}_{-0.12}$  \\ 
127                         &  3560 $\pm$145             &      0.34$^{+0.41}_{-0.19}$   &   2.87$^{+5.76}_{-1.83}$    & 0.56$^{+0.17}_{-0.13}$ &  2.62$^{+2.87}_{-1.55}$     & 0.38$^{+0.09}_{-0.07}$  \\ 
130                         &  4205 $^{+385}_{-355}$     &      0.73$^{+0.87}_{-0.41}$   &   7.88$^{+16.91}_{-5.44}$   & 1.19$^{+0.18}_{-0.27}$ &  4.39$^{+12.07}_{-2.68}$    & 0.95$^{+0.20}_{-0.38}$  \\ 
131                         &  8720 $^{+800}_{-695}$     &     13.75$^{+16.42}_{-7.74}$  &                             &                        &  6.06$^{+3.67}_{-6.00}$     & 1.94$^{+0.42}_{-0.18}$  \\ 
138\tablenotemark{\dagger}  &  3850 $^{+500}_{-435}$     &      3.73$^{+4.46}_{-2.10}$   &                             &                        &                             &                         \\ 
139                         &  8460 $^{+1060}_{-880}$    &     17.96$^{+21.44}_{-10.10}$ &                             &                        &  8.88$^{+40.03}_{-8.80}$    & 2.01$^{+0.51}_{-0.27}$  \\ 
141                         &  8720 $^{+800}_{-695}$     &     38.80$^{+46.31}_{-21.83}$ &                             &                        &  4.51$^{+5.16}_{-1.84}$     & 2.51$^{+0.54}_{-0.51}$  \\ 
142                         &  3270 $^{+290}_{-280}$     &      0.07$^{+0.09}_{-0.04}$   &   5.70$^{+19.62}_{-4.92}$   & 0.24$^{+0.26}_{-0.14}$ &  6.06$^{+8.12}_{-2.96}$     & 0.21$^{+0.13}_{-0.10}$  \\ 
145                         &  5845 $^{+185}_{-75}$      &      3.15$^{+3.77}_{-1.77}$   & 180.47$^{+511.90}_{-178.57}$& 2.01$^{+1.34}_{-0.81}$ & 16.48$^{+44.83}_{-7.78}$    & 1.33$^{+0.36}_{-0.17}$  \\ 
146                         &  3270 $^{+290}_{-280}$     &      0.27$^{+0.33}_{-0.15}$   &   1.05$^{+2.82}_{-1.05}$    & 0.28$^{+0.25}_{-0.13}$ &  2.35$^{+1.75}_{-59.44}$    & 0.25$^{+0.12}_{-0.04}$  \\ 
148                         &  4060 $^{+145}_{-210}$     &      0.52$^{+0.62}_{-0.29}$   &   8.54$^{+18.89}_{-5.82}$   & 1.06$^{+0.13}_{-0.22}$ &  4.72$^{+13.78}_{-2.62}$    & 0.79$^{+0.16}_{-0.22}$  \\ 
149                         &  3850 $^{+210}_{-145}$     &      0.14$^{+0.17}_{-0.08}$   &  29.53$^{+66.82}_{-19.81}$  & 0.74$^{+0.13}_{-0.17}$ & 17.36$^{+48.16}_{-12.23}$   & 0.59$^{+0.10}_{-0.13}$  \\ 
150                         &  4350 $^{+550}_{-500}$     &      1.13$^{+1.35}_{-0.63}$   &   6.04$^{+13.30}_{-4.31}$   & 1.37$^{+0.18}_{-0.35}$ &  3.46$^{+9.21}_{-2.69}$     & 1.13$^{+0.12}_{-0.57}$  \\
\enddata
\tablenotetext{a}{post-AGB stars, confirmed with Spitzer IRS spectra}
\tablenotetext{b}{Objects without 2MASS $J$, $H$ and $K$ magnitudes}
\tablenotetext{\dagger}{background objects}
\tablecomments{Subsript B corresponds to values derived from the \citet{BA98} models, while S corresponds to values derived from \citet{SI00}}
\end{deluxetable}

\begin{deluxetable}{l c c c c c }
\tabletypesize{\footnotesize}
\tablecolumns{6}
\tablewidth{0pt}
\tablecaption{H$\alpha$ equivalent widths and line widths, and mass accretion rates for YSOs in Serpens \label{t_alpha}} 
\tablehead{\colhead{\#}                 & 
           \colhead{EW[H$\alpha$]\tablenotemark{a}}  & 
           \colhead{CTTS}               & 
           \colhead{H$\alpha$[10\%]}    & 
           \colhead{CTTS}               & 
           \colhead{log $\dot{M}_{ac}$} \\
           \colhead{}                   & 
           \colhead{(6563 \AA{})}       &
           \colhead{(EW[H$\alpha$])}    & 
           \colhead{(km/s)}             & 
           \colhead{(H$\alpha$[10\%])}  & 
           \colhead{(M$_\odot$/yr)}      }
\startdata
     1 &        4.6        &    $\surd$      &  469 $\pm$  88  &     $\surd$       &  -8.34 $\pm$ 0.96  \\
     3 &       61.5        &    $\surd$      &  539 $\pm$  59  &     $\surd$       &  -7.66 $\pm$ 0.75  \\ 
     6 &       23.9        &    $\surd$      &  805 $\pm$ 110  &     $\surd$       &  -5.08 $\pm$ 1.24  \\ 
     7 &       55.5        &    $\surd$      &  464 $\pm$  87  &     $\surd$       &  -8.39 $\pm$ 0.95  \\ 
    14 &       23.9        &    $\surd$      &  448 $\pm$ 112  &     $\surd$       &  -8.54 $\pm$ 1.17  \\ 
    16 &        --         &                 &                 &                   &                    \\
    29 &      104.9        &    $\surd$      &  333 $\pm$  84  &     $\surd$       &  -9.66 $\pm$ 0.90  \\  
    30 &       11.2        &    $\surd$      &  284 $\pm$  61  &     $\surd$       & -10.13 $\pm$ 0.69  \\ 
    36 &      128.1        &    $\surd$      &  340 $\pm$ 292  &     $\surd$       &  -9.59 $\pm$ 2.86  \\ 
    40 &        --         &                 &                 &                   &                    \\
    41 &       -0.7        &                 &                 &                   &                    \\ 
    43 &       19.1        &    $\surd$      &  241 $\pm$  94  &                   & -10.55 $\pm$ 0.97  \\ 
    44 &        --         &                 &                 &                   &                    \\ 
    47 &        --         &                 &                 &                   &                    \\ 
    48 &      332.3        &    $\surd$      &  762 $\pm$ 127  &     $\surd$       &  -5.50 $\pm$ 1.37  \\ 
    52 &       -2.2        &                 &                 &                   &                    \\ 
    53 &       92.7        &    $\surd$      &  529 $\pm$ 112  &     $\surd$       &  -7.76 $\pm$ 1.19  \\  
    55 &       11.2        &    $\surd$      &  413 $\pm$  70  &     $\surd$       &  -8.88 $\pm$ 0.80  \\
    58 &       71.6        &    $\surd$      &  507 $\pm$  63  &     $\surd$       &  -7.97 $\pm$ 0.77  \\
    60 &       44.8        &    $\surd$      &  545 $\pm$ 104  &     $\surd$       &  -7.60 $\pm$ 1.12  \\ 
    61 &       28.5        &    $\surd$      &  311 $\pm$  57  &     $\surd$       &  -9.87 $\pm$ 0.67  \\ 
    62 &        --         &                 &                 &                   &                    \\ 
    66 &      127.4        &    $\surd$      &  966 $\pm$ 138  &     $\surd$       &  -3.52 $\pm$ 1.53  \\ 
    70 &      -15.1        &                 &                 &                   &                    \\ 
    71 &       68.1        &    $\surd$      &  490 $\pm$  84  &     $\surd$       &  -8.14 $\pm$ 0.93  \\ 
    76 &       -9.7        &                 &                 &                   &                    \\ 
    82 &       25.2        &    $\surd$      &  656 $\pm$ 123  &     $\surd$       &  -6.53 $\pm$ 1.31  \\ 
    87 &       --          &                 &                 &                   &                    \\ 
    88 &       --          &                 &                 &                   &                    \\
    89 &       22.2        &    $\surd$      &  532 $\pm$ 174  &     $\surd$       &  -7.73 $\pm$ 1.75  \\ 
    91 &      108.9        &    $\surd$      &  549 $\pm$  65  &     $\surd$       &  -7.56 $\pm$ 0.80  \\ 
    92 &       24.8        &    $\surd$      &  478 $\pm$ 138  &     $\surd$       &  -8.25 $\pm$ 1.41  \\ 
    96 &       22.0        &    $\surd$      &  704 $\pm$ 168  &     $\surd$       &  -6.06 $\pm$ 1.73  \\ 
    97 &        --         &                 &                 &                   &                    \\ 
    98 &       -7.8        &                 &                 &                   &                    \\ 
   106 &       15.8        &    $\surd$      &  175 $\pm$ 186  &                   & -11.19 $\pm$ 1.83  \\ 
   113 &       10.9        &    $\surd$      &  273 $\pm$  73  &     $\surd$       & -10.24 $\pm$ 0.79  \\ 
   114 &        --         &                 &                 &                   &                    \\ 
   115 &        --         &                 &                 &                   &                    \\ 
   117 &       91.9        &    $\surd$      &  559 $\pm$ 152  &     $\surd$       &  -7.47 $\pm$ 1.55  \\ 
   119 &        8.8        &                 &  639 $\pm$  94  &     $\surd$       &  -6.69 $\pm$ 1.06  \\ 
   120 &      -11.0        &                 &                 &                   &                    \\ 
   121 &        --         &                 &                 &                   &                    \\ 
   122 &       37.1        &    $\surd$      &  444 $\pm$  76  &     $\surd$       &  -8.58 $\pm$ 0.85  \\ 
   123 &        --         &                 &                 &                   &                    \\ 
   124 &       34.7        &    $\surd$      &  225 $\pm$  64  &                   & -10.71 $\pm$ 0.71  \\ 
   125 &       49.0        &    $\surd$      &  618 $\pm$  95  &     $\surd$       &  -6.89 $\pm$ 1.06  \\ 
   127 &        5.6        &                 &  460 $\pm$ 213  &     $\surd$       &  -8.43 $\pm$ 2.11  \\ 
   130 &      123.2        &    $\surd$      &  590 $\pm$ 114  &     $\surd$       &  -7.17 $\pm$ 1.22  \\ 
   131 &       -9.6        &                 &                 &                   &                    \\ 
   139 &       -6.4        &                 &                 &                   &                    \\ 
   141 &        9.0        &    $\surd$      &  289 $\pm$ 100  &     $\surd$       & -10.09 $\pm$ 1.03  \\ 
   142 &      116.6        &    $\surd$      &  367 $\pm$  75  &     $\surd$       &  -9.33 $\pm$ 0.83  \\ 
   145 &       -1.5        &                 &                 &                   &                    \\ 
   146 &       23.4        &    $\surd$      &  318 $\pm$ 102  &     $\surd$       &  -9.80 $\pm$ 1.06  \\ 
   148 &       18.6        &    $\surd$      &  598 $\pm$ 122  &     $\surd$       &  -7.09 $\pm$ 1.29  \\ 
   149 &      226.1        &    $\surd$      &  502 $\pm$  36  &     $\surd$       &  -8.02 $\pm$ 0.58  \\ 
   150 &        3.4        &    $\surd$      &  213 $\pm$ 198  &                   & -10.82 $\pm$ 1.95  \\
\enddata
\tablenotetext{a}{Positive values denote emission}
\end{deluxetable}

\end{document}